\documentclass[aps,prd,twocolumn,superscriptaddress,amsfonts,amssymb,amsmath,eqsecnum,nofootinbib,floatfix]{revtex4}

\usepackage{graphicx,color,ulem}
\usepackage{subcaption}
\usepackage{algorithm}
\usepackage{algorithmicx}
\usepackage[noend]{algpseudocode}
\definecolor{darkblue}{rgb}{0,0,0.7}
\definecolor{darkred}{rgb}{0.7,0,0}
\usepackage[unicode, colorlinks, citecolor=darkblue, linkcolor=darkred, urlcolor=blue]{hyperref}

\newcommand{\beq}{\begin{equation}}
\newcommand{\eeq}{\end{equation}}
\newcommand{\bea}{\begin{eqnarray}} 
\newcommand{\eea}{\end{eqnarray}}

\algdef{SE}[DOWHILE]{Do}{doWhile}{\algorithmicdo}[1]{\algorithmicwhile\ #1}%

\allowdisplaybreaks

\definecolor{dgreen}{rgb}{.3,.7,.3}

\begin{document}
	\date{\today}
	
	\title{Spinning black holes as cosmic string factories}
	
	\author{Hengrui Xing}
	\affiliation{Department of Physics, Columbia University, New York, NY 10027, USA}
	\author{Yuri Levin}
	
	
	\affiliation{Department of Physics, Columbia University, New York, NY 10027, USA}
	\affiliation{ Center for Computational Astrophysics, Flatiron Institute, New York, NY 10010, USA}
	\affiliation{School of Physics and Astronomy, Monash Center for Astrophysics, Monash University, Clayton, VIC 3800, Australia}
	
	\author{Andrei Gruzinov}
	\affiliation{CCPP, Department of Physics, New York University,
		New York, NY 10001}
	
	\author{Alexander Vilenkin}
	\affiliation{Institute of Cosmology, Department of Physics and Astronomy,
		Tufts University, Medford, MA 02155}

	\begin{abstract} \noindent
		We consider the evolution of a cosmic string loop that is captured by a much more massive and much more compact black hole. We show that after several reconnections that produce ejections of smaller loops, the loop that remains bound to the black hole  moves on a nearly-periodic non-self-intersecting trajectory, ``the orbit". The orbit   evolves due to an energy
		and angular momentum exchange between the loop and the spinning black hole. We show that such evolution is mathematically equivalent to a certain continuous deformation of an auxiliary closed curve in a 3-dimensional space; for zero black-hole spin this deformation is {\it curve-shortening} that has been extensively studied by mathematicians as a prominent example of one-dimensional geometric flows.
		The evolution features competing effects of loop growth by the superradiant extraction of the black-hole spin energy, and loop decay by the friction of the moving string against the horizon. Self-intersection of an auxiliary curve may be a common occurrence, which correspond to a capture by the black hole of a new string segment and thus an addition of a new captured loop. Possible asymptotic states of such evolution are explored and are shown to be strong emitters of gravitational waves. Whether reconnections prevent reaching the asymptotic states remains to be explored.
Additionally, the orbit's shape also evolves due to 1.~an emission of gravitational waves, and 2.~a finite mass of the black hole, which leads to the recoil that secularly 
        changes the orbit and likely leads to self-intersections.
		
		We argue that for a significant range of the dimensionless tension $\mu$, 
		string loops are captured by 
		supermassive black holes at the centers of galaxies. This strongly motivates further study of interaction
		between string loops and black holes, especially the influence of this process on the black hole
		spindown and on the production of gravitational waves by strings created in galactic nuclei.
		We also discuss potential loop captures by primordial black holes.
	\end{abstract}
	%
	\pacs{} 
	\maketitle
	\section{Introduction}
	Black holes are fundamental objects in theoretical physics, and at the same time they are the subject of intense 
	study by much of modern astronomy and astrophysics. Cosmic strings do not share the same status with black holes. They arise naturally
	in theoretical physics as possible  remnants of a phase transition in the early Universe \citep{1976JPhA....9.1387K,2000csot.book.....V}. Fundamental strings of superstring theory can be formed at the end of brane inflation and can also play the role of cosmic strings \cite{2003PhLB..563....6J}. Moreover, the structure of spacetime
	generated by an undisturbed, straight cosmic string is very simple: the geometry of a plane perpendicular to the
	string is that of a cone, with the deficit angle proportional to the string tension. The motion of a string is specified by minimizing its Nambu-Goto action,  which equals the area of the surface swept in space-time by the string's trajectory. Therefore, a string is also a fundamental relativistic object in theoretical physics. However, there is currently no observational evidence for the existence of cosmic strings in our Universe.
	Still, if they do exist they may produce a number of potentially detectable phenomena, notably a stochastic background of gravitational waves from oscillating cosmic string loops \citep{1981PhLB..107...47V,2005PhRvD..71f3510D}.
	
	In this paper we investigate the hypothetical interaction of a cosmic string loop with a black hole. In particular, we are interested in what happens to the loop once a small part of it gets captured by the black hole. Our motivation for studying this is two-fold. 
	Firstly, as we show later, for reasonable values of $\mu$, string loops are expected to be captured by supermassive black holes in galactic nuclei. The captures will also take place if both a cosmic string network and a multitude of primordial black holes formed in the early Universe \cite{2018JCAP...11..008V}. 
	Secondly, non-trivial interaction between two fundamental relativistic objects is a good problem in its own right, and it might become relevant in a context that we cannot foresee. 
	
	The interaction between cosmic strings and black holes has been investigated in a number of previous studies. Gravitational capture and the scattering of an initially straight string by a black hole have been studied in \citep{1988NuPhB.298..693L,1998IJMPD...7..957D,1999CQGra..16.2403D,2003CQGra..20.1303S,2007IJMPD..16.1311D}. Stationary horizon-crossing cosmic string solutions in Kerr metric were found in \cite{1996PhRvD..54.5093F,2018PhRvD..98f4021I}, and it was demonstrated that the stationary string can extract angular momentum from the black hole. We will be guided by the results from these papers in our exploration of the dynamics of a string loop captured by a black hole. The plan of our 
	paper is as follows. In Section 2, we discuss the dynamics of a loop bound to a black hole, in the limit where the loop mass is zero, and show that such motion is specified by a stationary 3-dimensional closed auxiliary curve. We demonstrate through numerical experiments that after several self-intersections and reconnections, a fraction of the loop's
	original length remains bound to the black hole and is moving on a stable non-self-intersecting trajectory. In Section
	3 we relax the assumption of zero-loop-mass, and show that the loop trajectory evolves secularly on a  timescale $\sim (M/m)P$, where $M$ and $m$ are the masses of the black hole and the string, and $P$ is the period of motion of the bound loop. In the absence of other effects this evolution would lead to a continuous chain of  physical self-intersections, that, if accompanied by efficient reconnections,  deplete the loop length approximately exponentially with time. In Section 4  we consider the exchange of energy and angular momentum between the string and the spinning and non-spinning black hole, and show that this leads to horizon friction,
	superradiance of tension waves and the spindown of the black hole.
	In Section 5 we develop a formalism that allows one to model the change in the loop shape,
	by showing that it corresponds to easily modeled
	 deformation of an auxiliary curve. We find and explore the late-time asymptotic states of the captured loops, which 
	 are shown to be particularly strong emitters of gravitational waves. In Section 6 we outline the evolutionary scenarios for loops captured by both spinning and non-spinning black holes, and we argue that the spinning black holes might be string factories that are converting their spin energy into string length. In Section 7
	we estimate the rate of loop captures by supermassive black holes in galactic nuclei as well as by primordial black holes, both of which are shown to be potentially significant. We argue qualitatively that 
	black-hole string factories in galactic nuclei may be prolific sources of gravitational waves which strongly motivates further study. In Section 8 we conclude.
	
	\section{Motion of an infinitely light string loop}
	\subsection{General solutions}
	In this work we focus on the case where the invariant length of a string loop is much greater than the black-hole gravitational radius, $L\gg R=GM/c^2$, but at the same time its mass is much smaller than that of the black hole, $G\mu L/c^4\ll R$; here $\mu$ is the string tension
	force, $L=mc^2/\mu$ is the invariant length of the string, and $m$ is the mass of the string.  From here on we will be using geometric units with ${\bf G=c=1}$; in these units  $\mu$ is dimensionless. 
	
	The  best observational bound on the string  tension $\mu$ is based on the lack of detection of gravitational waves from freely oscillating string loops by Pulsar Timing Arrays. A somewhat model-dependent constraint $\mu<1.5\times 10^{-11}$ has been obtained in \cite{2018PhLB..778..392B}. This small value of the dimensionless tension allows a large range of possible string length that satisfies our constraints,  $1\ll L/R\ll (\mu)^{-1}$. 
	
	In this section we focus on the kinematics of string motion and we will neglect its influence on the black hole position. In this approximation of an infinitely light string, we model the influence of the black hole by rigidly fixing a point on the loop in space. The rest 
	of the loop is assumed to move in Minkowski space; we neglect the general-relativistic character of the string motion
	through the curved spacetime at distances $\sim R$ from the black hole, since $R\ll L$. 
	
	The motion of the free part of the string (i.e., all of it except for the pinned point) is given by
	\begin{equation}
		{\bf r}(\sigma, t)={1\over 2}\left[ {\bf a}(\sigma-t)+{\bf b}(\sigma+t)\right].
		\label{eq1}
	\end{equation}
	Here $\sigma$ is the invariant length coordinate marking points along the string,  ${\bf r}=(x,y,z)$ is the position of a string point $\sigma$ at time $t$, and ${\bf a}$ and ${\bf b}$ are vector functions of a single variable, such that $\left|{\bf a}^{\prime}\right|=\left|{\bf b}^{\prime}\right|=1$; see \cite{2000csot.book.....V} for derivation. We shall take $\sigma=0$ and $\sigma=L$ at the black hole location ${\bf r}=0$. The boundary conditions ${\bf r}(0,t)={\bf r}(L,t)=0$ lead to the following constraints:
	\begin{eqnarray}
		{\bf a}(\eta)=-{\bf b}(-\eta),\nonumber\\
		{\bf a}(\eta)={\bf a}(\eta+2L).\label{newconstraints}
	\end{eqnarray}
	Here $\eta$ is a point on a real axis, $-\infty<\eta<\infty$. The general solution for the pinned loop is therefore given by
	\begin{equation}
		{\bf r}(\sigma, t)={1\over 2}[{\bf a}(\sigma-t)-{\bf a}(-\sigma-t)],\label{gensol}
	\end{equation}
	with the requirement that ${\bf a}$ is periodic with the period of $2L$.
	Formally ${\bf r}(\sigma, t)$ is defined for all real values of $\sigma$ and $t$, but the physical loop corresponds to the values of $0\le\sigma\le L$. Extending $\sigma$ to the interval between $L$ and $2L$ would 
	produce a ``ghost'' loop obtained from the original loop by reflection with respect to the origin, since the solution observes the symmetry 
	\begin{equation}
		{\bf r}(\sigma, t)=-{\bf r}(2L-\sigma,t).
	\end{equation}
	In other words, a loop pinned at a point can be considered as half of a free loop that self-intersects at the origin and has a reflection symmetry with respect to the origin.
	
	\begin{figure}[t]
		\centering
		\includegraphics[width=.51\textwidth]{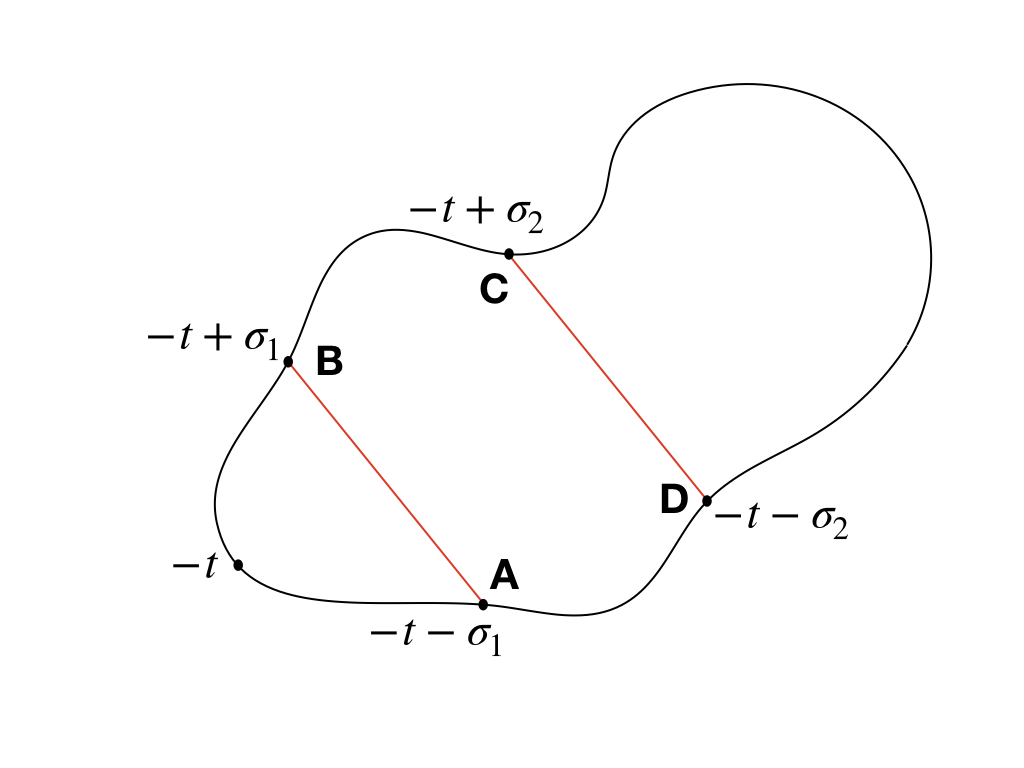}
		\caption{Auxiliary closed curve ${\bf a}(\sigma)$; a position vector ${\bf r}(\sigma,t)$ corresponds to a directed chord connecting points on the curve with coordinates $-t-\sigma$ and $-t+\sigma$. The energy and angular momentum of the loop equal $\mu$ multiplied by the half-length and half of the directed area of the auxiliary curve, respectively. A pair of parallel chords $AB$ and $DC$ mark a self-intersection of the loop iff their lengths are equal and if the path lengths along the curve from $A$ to $D$ and from $B$ to $C$ are equal. After reconnection and ejection of the newly formed loop, the remaining loop is described by a new closed curve obtained from the old one by gluing the $2$ chords.}
		\label{figstring1}
	\end{figure}

	\subsection{Self-intersections}
	Self-intersections are important because they can lead to reconnections and ejections of daughter loops,
	which deplete the loop bound to the black hole. The reconnection probability $p$ for a cosmic string that is a solution of a classical gauge field theory is close to unity \cite{2000csot.book.....V}, unless the segments collide at ultra-relativistic velocities \cite{2006PhRvD..74l1701A}. On the other hand, the reconnection probability for cosmic super-strings may be smaller than unity by orders of magnitude, plausibly as small as $10^{-3}$ in some superstring models \cite{2005JHEP...10..013J}. In this work we keep an open mind about the value of $p$, but it is clear that reconnections are very important for the loop evolution.
	
	\paragraph{Geometric interpretation.}
	Equation (\ref{gensol}) provides us with a geometrical interpretation of the loop's dynamics.
	The loop trajectory is completely specified by an auxiliary closed curve of length $2L$ in $3$-dimensional space, ${\bf a}(\sigma)$, where $\sigma$ is the cyclic coordinate that marks the length along the curve. The position vector of ${\bf r}(\sigma,t)$ equals half of the directed chord connecting $2$ points
	on the curve ${\bf a}$. The energy $E$ and angular momentum ${\bf \Lambda}$ of the loop have a simple  geometrical interpretation in terms of the length $L_a=\oint_0^{2L}d\sigma$ and the directed area ${\bf S}_a=0.5\oint_0^{2L}{\bf a} \times {\bf a}^{\prime} d\sigma$ of the auxiliary curve:
	\begin{eqnarray}
	  E&=&{1\over 2}\mu L_a.\nonumber\\
	  {\bf \Lambda}&=&-{1\over 2} \mu {\bf S}_a.
	  \label{EJloop}
	\end{eqnarray}
	
	From Eq.~(\ref{gensol}) one can see that a self-intersection of the loop corresponds to a pair of chords $AB$ and $DC$ that
	are parallel to each other, equal in size, and such that the length along the curve between the points $A$ and $D$ equals that between $B$ and $C$; see Figure \ref{figstring1}.  After the self-intersection takes place, if the string reconnects, the new bound loop corresponds to the new closed curve that is obtained by throwing away the segments $\overline{AD}$ and $\overline{BC}$ of the old curve and gluing $A$ to $D$ and $B$ to $C$.
	
	It is quite straightforward to find examples of the closed curves ${\bf a}(\sigma)$ that do not allow for self-intersections. For a concrete example, consider\newline\newline
	${\bf a}=\sigma {\bf e}_x$ for $0<\sigma<1$,\newline
	${\bf a}={\bf e}_x+(\sigma-1) {\bf e}_y$ for $1<\sigma<2$,\newline
	${\bf a}={\bf e}_x+{\bf e}_y+(\sigma-2) {\bf e}_z$ for $2<\sigma<3$,\newline
	${\bf a}=[1-(\sigma-3)/\sqrt{3}]\left({\bf e}_x+{\bf e}_y+{\bf e}_z\right)$ for $3<\sigma<3+\sqrt{3}$.\newline\newline
	In other words, ${\bf a}(\sigma)$ follows a closed curve via $3$ equal steps along  x, then y, then z, and then back to the origin. One can easily inspect that this quadrilateral does not have a pair of chords that satisfy the conditions described above, and therefore the string loop corresponding to this closed curve does not have self-intersections. A little more work is required to show that if ${\bf a}(\sigma)$ is a general, non-flat quadrilateral, then the corresponding loop is self-intersecting only if the quadrilateral's side lengths are fine-tuned.
	
	\begin{figure}[t]
		\centering
		\includegraphics[width=.51\textwidth]{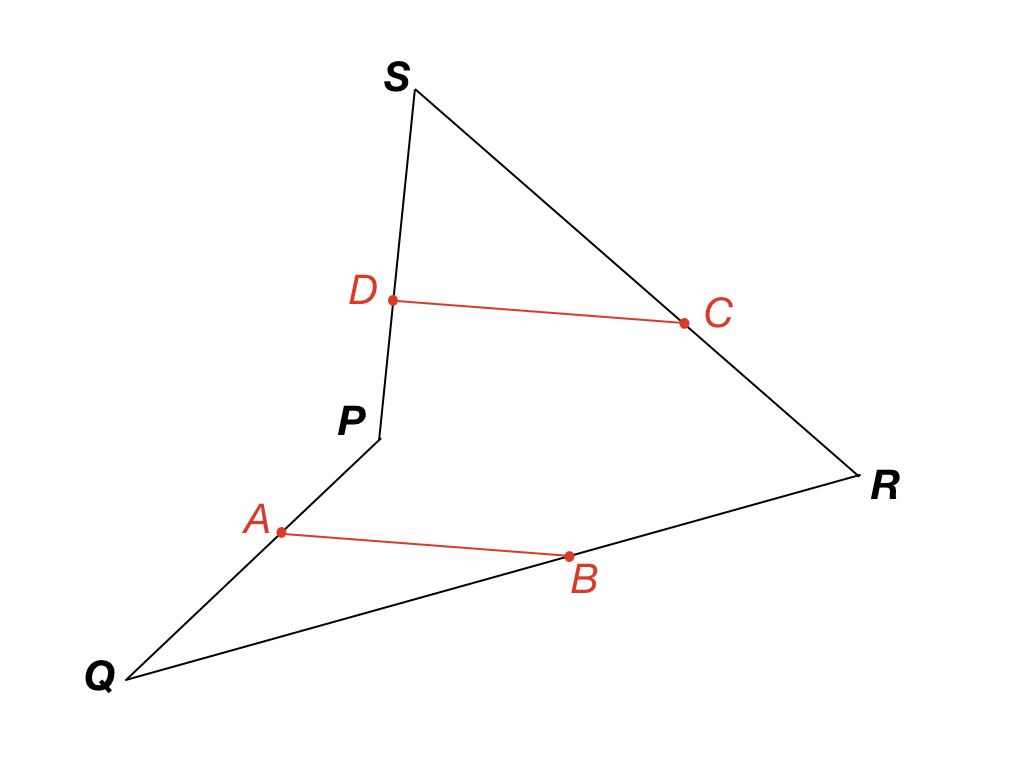}
		\caption{In this example, ${\bf a}(\sigma)$ is a non-flat quadrilateral $PQRS$. The self-intersection, if it exists, is marked by equal chords $AB$ and $DC$ that must be parallel
			to one of the diagonals, here $PR$. As explained in the text,  self-intersections exists if and only if $|PQ|+|PS|=|RQ|+|RS|$.}
		\label{figstring2}
	\end{figure}

	\paragraph{Quadrilaterals.} Consider a general non-flat quadrilateral $PQRS$, and assume that there exists a parallelogram $ABCD$ with vertices lying on the quadrilateral's sides; see Figure \ref{figstring2}. First assume that two vertices, say $A$ and $B$, lie on the same side of the quadrilateral, say $PQ$. A plane that 
	passes through $PQ$, either contains $QR$, or contains $PS$, or intersects $RS$ at a single point, or does not intersect the rest of the quadrilateral. $C$ and $D$ must lie in one such plane, but in that case it is clear that $CD$ cannot be parallel to $AB$. Therefore $ABCD$ cannot have two of
	its vertices  on the quadrilateral sides, and instead it must have one vertex located on each of the quadrilateral's sides. 
	
	From this it also follows that neither $AB$ nor $CD$ can have their vertices located on the opposite sides of the quadrilateral. To show this, suppose (say) $AB$ was so located. $C$ and $D$ must be located on the same side of the curve ${\bf a}(\sigma)$ relative to $AB$, in a sense that one could move along the curve from $C$ to $D$ without encountering $A$ or $B$. But this implies that at least $2$ of $A$,$B$,$C$, and $D$ would have to be on the same quadrilateral side.
	
	Therefore the only possible way for parallel chords $AB$ and $DC$ to be accommodated is for their ends to lie on the neighbouring sides of the quadrilateral, e.g.~ for $A$ on $PQ$, $B$ on
	$QR$, $C$ on $RS$, and $D$ on $SP$.  Since $AB$ and $DC$ are parallel to each other, they should both be parallel to $PR$.  The fact that $|AB|=|DC|$ implies that $|PA|/|PQ|=|PD|/|PS|$.
	With these choices, the remaining constraint $|PA|+|PD|=|RB|+|RC|$ is satisfied if and only if $|PS|+|PQ|=|RS|+|RQ|$. Clearly the set of quadrilaterals that satisfy this constraint has measure zero relative to the set of general non-flat quadrilaterals.
	
	
	\subsection{Numerical experiments} 
	\label{sec:num}
	The fact that a general quadrilateral auxiliary curve ${\bf a}(\sigma)$ corresponds to a non-self-intersecting loop  makes it plausible that such loops are pretty common, and that
	a general-shape  loop will settle into a non-self-intersecting configuration after several reconnections. We did not manage to find a rigorous mathematical proof to this statement. Instead we carried out numerical experiments with assumed reconnection probability $p=1$, that showed that this is indeed  what happens to loops with initially arbitrary shapes.
	
	The reader uninterested in details is urged to accept this statement on faith and skip the rest of this subsection. Methodologically, we are interested in how the number of reconnections and the length of the final loop depends on the complexity of the initial loop. Our task is then to first, introduce some way of initializing loops of variable  complexity, and second, to develop a reliable and efficient algorithm that searches for self-intersections of a moving loop.
	
		\begin{figure*}
		\centering
		\begin{subfigure}{0.85\textwidth}
			\includegraphics[width=\linewidth]{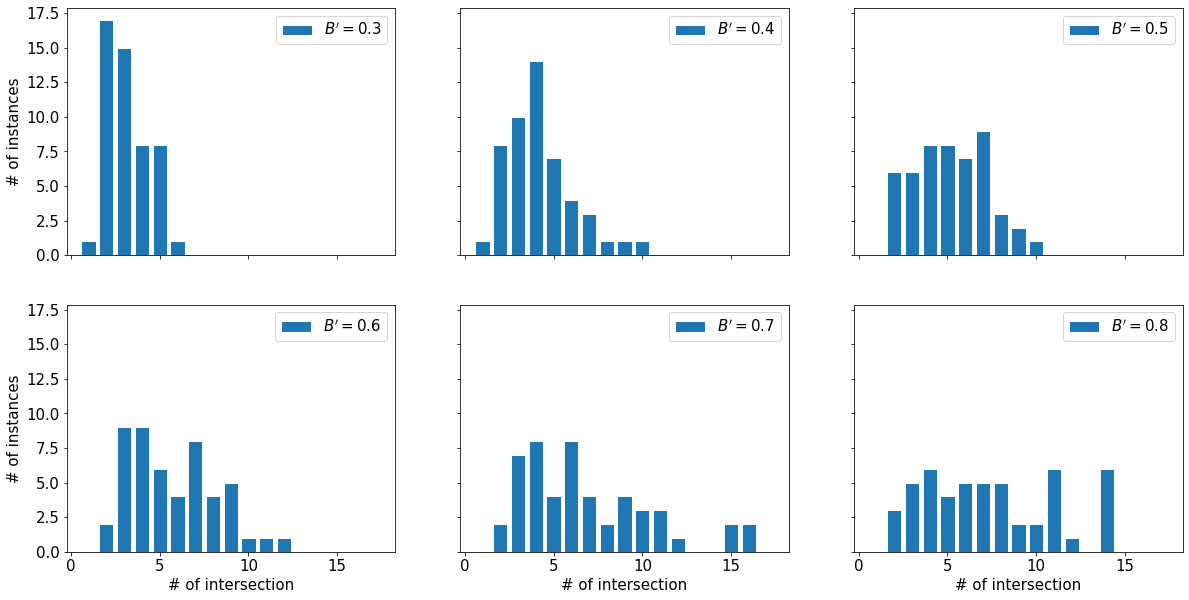}  
			\caption{Number of self-intersections before string stabilizes.}
			\label{fig:ni}
		\end{subfigure}
		\begin{subfigure}{0.85\textwidth}
			\includegraphics[width=\linewidth]{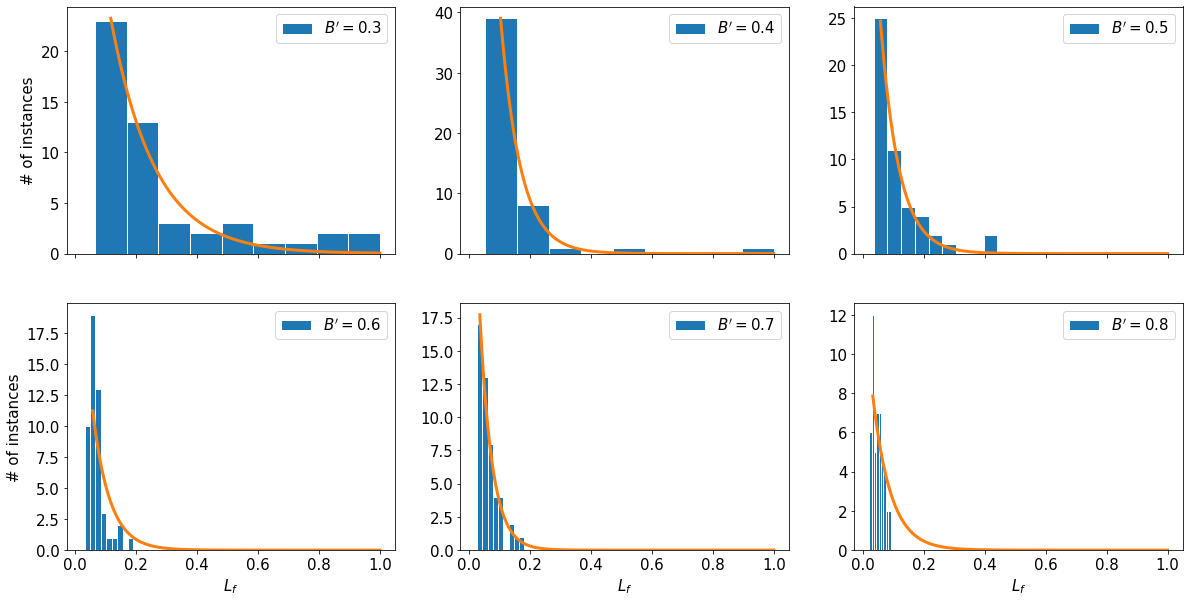}  
			\caption{Final length after string stabilizes. Orange curve shows fitted exponential distribution.}
			\label{fig:lf}
		\end{subfigure}
	
	\begin{subfigure}{0.4\textwidth}
		\includegraphics[width=\linewidth]{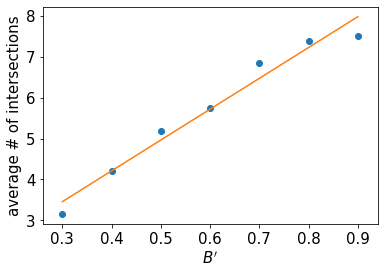}  
		\caption{Linear relationship between average number of self-intersections and relative curvature bound. $R^2=97\%$.}
		\label{fig:int_line}
	\end{subfigure}
	\quad
	\begin{subfigure}{0.4\textwidth}
\includegraphics[width=\linewidth]{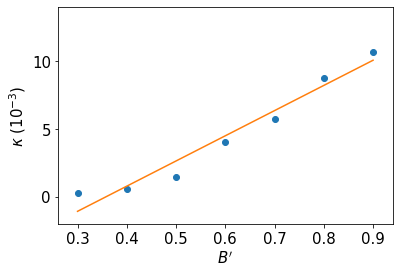}  
\caption{Linear relationship between exponential decay constant for $L_f$ and relative curvature bound. $R^2=96\%$.}
\label{fig:kappa_line}
\end{subfigure}
		\caption{Numerical experiment results for randomly initialized string loops attached to an infinitely massive BH.}
		\label{fig:stat}
	\end{figure*}

	We construct the initial loop as follows.
	We discretize the ${\bf a}$ vector function into $2N$ connected line segments. Each segment is defined by its length $L_i$ and direction unit vectors ${\bf a}^{\prime}_i$ for $i=1,\dots,2N$. For a point on ${\bf a}$ that falls into the $j$th segment, we have:
	\begin{equation}
		{\bf a}(\eta) = {\bf a}(0)+\sum_{i<j}L_i{\bf a}^{\prime}_i + \left(\eta-\sum_{i<j}L_i\right){\bf a}^{\prime}_i.\label{approx}
	\end{equation}
	Since $	{\bf a}$ should be periodic in $2L$, we require that $\sum_{i=1}^{2N}L_i=2L$ and $\sum_{i=1}^{2N}L_i{\bf a}^{\prime}_i={\bf 0}$. We also note that 
	${\bf r}$ does not change if we add a constant vector to ${\bf a}$. Therefore, we 
	assign 
	$ {\bf a}(0)={\bf 0}$ in the simulation.
	
	To make the curve random, the direction vectors ${\bf a}^{\prime}_i$ perform a random walk 
	on a unit sphere; the larger is the step of the random walk, the greater is the complexity of the loop. 
	For the self-intersection search, it is convenient for all segment lengths to be equal $L_i=L/N$. 
	Our algorithm 
	 generates the set of random ${\bf a}^{\prime}_i$ which satisfies:
	\begin{enumerate}
		\item $|{\bf a}^{\prime}_i|=1$ (on unit sphere)
		\item $\sum {\bf a}^{\prime}_i=0$ (periodicity)
		\item $|{\bf a}^{\prime}_{i+1}-{\bf a}^{\prime}_i|\le B^{\prime}=2\sin(\theta_m/2)$ (smoothness)
	\end{enumerate}
	Here $\theta_m$ is the maximum angle between the neighbouring segments; it is assumed that $1/N\ll\theta_m<1$.  
	Our  iterative procedure is described in Appendix \ref{app:gen}. 
	
	We also design an algorithm for detecting self-intersections and for implementing reconnections that follow. An intersection occurs when, for some $0\le\sigma_i\ne \sigma_j<L$, ${\bf r}(\sigma_i, t)={\bf r}(\sigma_j, t)$. We can use Eq. (\ref{gensol}) and Eq. (\ref{approx}) to formulate a linear equation and solve for $\sigma_i$, $\sigma_j$ and $t$. After intersection, the relevant portion of ${\bf a}$ will break off from the original function. In practice, since ${\bf a}$ is periodic in $2L$, we search for self-intersection in the time range $0\le t < 2L$. The simulation method is described in Appendix \ref{app:det}. The algorithm was tested  (a) by running it on loops specified by ${\bf a}(\sigma)$ that were general quadrilaterals, to check that it does not find spurious intersections, and (b) by performing resolution tests in segment lengths and timesteps.
	
	We simulate strings with $L=1$, $N=500$ and a range of different smoothness constraints. Fifty random initializations are used for each $B^{\prime}$. We find that there exists loops that are stable (i.e., non-self-intersecting). A randomly initialized loop will always, after a series of self-intersection and reconnection, reduce to a stable configuration. 
	
	Figure \ref{fig:ni} shows a histogram of the number of reconnections a string undergoes before it stabilizes. A larger $B^\prime$
	causes more complexity in the initial loop and leads to more self-intersections. As illustrated in Figure \ref{fig:int_line}, the average number of intersections scales approximately linearly with $B^\prime$. Figure \ref{fig:lf} shows the distribution of $L_f$, the final invariant length of the string after all self-intersections. The result approximately follows an exponential distribution 
	\begin{equation}
	   P(L_f)\sim {\kappa\over L}e^{-\kappa L_f/L} 
	\end{equation} 
	where $\kappa$ is the decay constant. A similar statistical result for free cosmic strings has been shown \citep{PhysRevD.40.277}. We further demonstrate in Figure \ref{fig:kappa_line} that there is an approximate linear relationship  $\kappa\sim 0.01 B^\prime$; note that this relation is only valid for $N=500$ and $\kappa>1$. 

	\section{String orbit  evolution due to the black hole's finite mass}
	The discussion of the previous section was based on the premise that the string loop is rigidly pinned by the black hole at a point in space. This ensures that the motion of the rest of the loop is strictly periodic and therefore non-self-intersecting solutions are stationary. However, once the black hole is allowed to move under the influence of the string tension, the motion of the string is no longer strictly periodic and one can expect 
	the string orbit to change secularly on a timescale 
	\begin{equation}
	    t_{\rm sec}\sim {M\over m}P=2R/\mu.
	\end{equation}
	We show this rigorously in this section, using perturbation theory with respect to the small
	parameter $m/M$ to derive the evolution of the string orbit with time.
	
	We begin by noting that even though the string and black hole are relativistic objects, the
	motion of the black hole is non-relativistic, with characteristic velocity $\sim m/M$. The 
	equation of motion of the black hole is given simply by the Newton's second law,
	\begin{equation}
      M\ddot{\bf r}_{\rm BH}=\mu({\bf n}_1+{\bf n}_2),	
      \label{newton}
	\end{equation}
	where ${\bf r}_{\rm BH}$ is the displacement of the black hole and ${\bf n}_1(t)$ and ${\bf n}_2(t)$ are the unit vectors at the 2 ends of the string that are pointing away from the black hole. The string satisfies the wave equation
	\begin{equation}
	    {\partial^2 {\bf r}\over \partial t^2}={\partial^2 {\bf r}\over \partial\sigma^2}
	    \label{wave}
	\end{equation}
	with the boundary conditions
	\begin{equation}
	    {\bf r}(\sigma_1,t)={\bf r}(L-\sigma_2,t)={\bf r}_{\rm BH}(t).
	\end{equation}
	The quantities $\sigma_1$ and $\sigma_2$ are not zero, because the string does work on the moving black hole and its invariant length changes. Each end of the string looses or gains invariant length, depending on the sign of work that the end segment is performing.
	We have
	\begin{eqnarray}
	 {\dot{\sigma}_1}&=&\dot{\bf r}_{\rm BH}\cdot {\bf n}_1,\nonumber\\
	 {\dot{\sigma}_2}&=&\dot{\bf r}_{\rm BH}\cdot {\bf n}_2.
	 \label{dotsigma}
	 \end{eqnarray}
	 Similar equations have been derived in \cite{2001NuPhB.595..402S} for the motion of massive monopoles attached to pairs of strings.
	 
	 
	 Equations (\ref{newton})--(\ref{dotsigma}) form a complete system that can be modeled numerically using standard methods. However, in direct brute-force methods, the smallness
	 of $r_{\rm BH}$ is not manifest. We therefore choose a different approach. We find the variable domain $[\sigma_1,L-\sigma_2]$ inconvenient. Note however that if $\sigma_1<0$ and $\sigma_2<0$, we can easily evaluate the string positions at the ends of the old $\sigma$-interval $[0,L]$:
	 \begin{equation}
	     {\bf r}_1(t)\equiv {\bf r}(0,t)={\bf r}_{\rm BH}(t)-\left({\partial {\bf r}\over \partial \sigma}\right)_{\sigma=0}\sigma_1,
	     \label{boundary1}
	 \end{equation}
	 and 
	  \begin{equation}
	     {\bf r}_2(t)\equiv {\bf r}(L,t)={\bf r}_{\rm BH}(t)+\left({\partial {\bf r}\over \partial \sigma}\right)_{\sigma=L}\sigma_2,
	     \label{boundary2}
	 \end{equation}
	 both correct to first order in $\sigma_1, \sigma_2\sim L(m/M)$. Given $r_{\rm BH}(t)$, $\sigma_1(t)$, and $\sigma_2(t)$, one could then use Eqs.~(\ref{boundary1}) and (\ref{boundary2}) as the boundary conditions
	 for the wave equation on the fixed $\sigma$-interval $[0,L]$. In fact, in this we are not limited to the negative values of $\sigma_1,\sigma_2$. In case they are positive, Equations
	 (\ref{boundary1}) and (\ref{boundary2}) represent simple incremental extensions of the string by lengths $\sigma_1,\sigma_2$ at each end. Therefore, to the first order in $m/M$, the motion of the string is represented by a wave equation (\ref{wave}) on the $\sigma$-interval $[0,L]$
	 with boundary conditions
	 \begin{eqnarray}
	     {\bf r}(0,t)&=&{\bf r}_1(t) \label{r01}\\
	      {\bf r}(L,t)&=&{\bf r}_2(t),\label{r02}
	 \end{eqnarray}
	 where the latter are given by Eqs.~(\ref{boundary1}) and (\ref{boundary2}). 
	 
	 The linear wave equation (\ref{wave}) can be solved for any given pair of
boundary values ${\bf r}_{1,2}$.  Perturbatively, ${\bf r}_{1,2}$ are
$2L$-periodic, and therefore they are in resonance with the unperturbed
(homogeneous) solutions of (3.3). This leads to a secular evolution of
the loop. The details follow.
	 
	 Recall that the general string trajectory is given by Eq.~(\ref{eq1})
\begin{equation}
{\bf r}(t,\sigma)=\frac{1}{2}[{\bf a}(\sigma-t)+{\bf b}(t+\sigma)]
\label{1}
\end{equation}
with ${\bf a}(\sigma)$ and ${\bf b}(\sigma)$ satisfying
\begin{equation}
|{\bf a}'|=|{\bf b}'|=1.
\end{equation}
We want to solve Eq.(\ref{1}) in the range $0<\sigma <L$, $0<t<\infty$ with the boundary conditions in Eqs.~(\ref{r01}) and (\ref{r02}).

The unperturbed solution (assuming the black-hole mass is infinite) was obtained in Section 2 and is given by 
\beq
{\bf r}_0(\sigma,t)=\frac{1}{2}[{\bf a}_0(\sigma-t)-{\bf a}_0(-\sigma-t)].
\label{xtsigma}
\eeq
with 
\beq
{\bf a}_0(\sigma+2L)={\bf a}_0(\sigma).
\label{2L}
\eeq
This solution is periodic with period $P=2L$.  If we substitute it in Eq.(\ref{newton}), we find that  in a properly chosen inertial frame, up to the linear order in $(m/M)$ the BH motion has the same period and has the property 
\beq
{\bf r}_{BH}(t+L)=-{\bf r}_{BH}(t).
\eeq
This also implies that ${\bf r}_{1,2}(t+L)=-{\bf r}_{1,2}(t)$.

Substituting the general solution (\ref{1}) in the boundary conditions (\ref{r01}) and (\ref{r02}), we obtain the relations
\beq
{\bf a}(-t)+{\bf b}(t)=2{\bf r}_1(t),
\label{r1}
\eeq
\beq
{\bf a}(L-t)+{\bf b}(t+L)=2{\bf r}_2(t).
\label{r2}
\eeq
Let us now use these relations to compare the string configuration at time $t+2L$ to that at time $t$.  We find
\begin{eqnarray}
{\bf b}(t+2L)-{\bf b}(t)&=&2[{\bf r}_2(t+L)-{\bf r}_1(t)]\nonumber\\
                        &=&-2[{\bf r}_2(t)+{\bf r}_1(t)].\label{bevol1}
\end{eqnarray}
Applying this relation iteratively $N$ times, we obtain
\beq
{\bf b}(t+2NL)={\bf b}(t)-2N[{\bf r}_1(t)+{\bf r}_2(t)].
\label{bn}
\eeq
Similarly, we find
\beq
{\bf a}(t+2NL)={\bf a}(t)+2N[{\bf r}_1(L-t)+{\bf r}_2(L-t)].
\label{an}
\eeq

Eqs.~(\ref{bn}) and (\ref{an}) describe the secular evolution of a string attached to a BH.  At lowest order in $m/M$ the string oscillates with a period $P=2L$, but due to boundary conditions it acquires a change in shape, whose amplitude grows linearly with the number of oscillations.  The string configuration changes significantly after $N\sim M/m$ oscillations.  Our approximation scheme breaks down at about the same time, but it can be extended by  adjusting the string solution and recomputing the cyclic function $r_{1,2}(t)$.

As the loop changes its shape, it may acquire an intersection after $\sim M/m$ oscillations. If the string reconnects as a result of the self-intersection, it will eject a daughter loop of length $\lesssim L$. The rate of length loss from a sequence of such intersections is given by
\begin{eqnarray}
    {dL\over dt}&\sim& -\left[\max\left({M\over m}{P\over L},~p^{-1}\right)\right]^{-1}\nonumber\\
    &\sim& -\left[\max\left({R\over \mu L}, ~p^{-1}\right)\right]^{-1},
\end{eqnarray}
where $p$ is the probability of reconnection when the two string segments cross each other. If $p\gg \mu L/R$ (we remind the reader that $p\sim 1$ for ordinary cosmic strings), then
 the loop shrinks approximately exponentially, on a timescale
\begin{equation}
    t_{\rm shrink}\sim R/\mu.
    \label{tshrink}
\end{equation}
As we will see in the next section, this is comparable to the timescale on which a spinning black hole transfers angular momentum to the string.

	\section{Angular momentum and energy exchange with the black hole}
	The loop motion takes place on a much longer timescale than the light-crossing time $R$ for the black hole, and we imagine that the loop is mostly
	smooth on the lengthscale $\sim R$. It is likely that a realistic loop has kinks. However, they will be quickly smoothed, 
	because the string distortions on  scales $\lesssim R$ get quickly absorbed by the black hole. We will thus assume that as viewed from the vicinity of the black hole, a string stretches towards the black hole horizon from a distance $\gg R$ on a nearly radial straight line, until it is curved by the
	dragging of the inertial frames near the black hole. The  asymptotic direction of the radial straight line changes slowly compared to $R$, so to first approximation the string is stationary (i.e., $t$-independent) in Boyer-Lindquist coordinates $(t, r,\theta, \phi)$ describing the Kerr metric of the black hole. Such stationary solutions have been explored in \cite{1989PhLB..224..255F}, \cite{1996PhRvD..54.5093F}, \cite{2018PhRvD..98f4021I}. It was shown that the string lies on a constant $\theta$ surface, and its shape is given by
	\begin{equation}
		\phi(r)=\phi_0+{a\over r_+-r_-}\log\left({r-r_-\over r- r_+}\right),
		\label{frolov}
	\end{equation}
	where $a$ is the Kerr parameter, and $r_{\pm}=R\pm\sqrt{R^2-a^2}$. When $r\gg R$, $\phi=\phi_0+a/r$, and the tension force along the string pulling away from the black hole is given by
	\begin{equation}
		{\bf F}\simeq \mu \left({\bf e}_r+r\sin\theta{\partial\phi\over \partial r}{\bf e}_{\phi}\right)=\mu \left[{\bf e}_r-{a\sin\theta\over r}{\bf e}_{\phi}\right],
	\end{equation}
	where  ${\bf e}_{r,\phi}$ are the unit vectors in $r,\phi$ directions. The torque applied to the black hole is given by
	\begin{equation}
		{\bf Q}={\bf r}\times {\bf F}=\mu a\sin\theta~{\bf e}_\theta.
		\label{torque}
	\end{equation}
	
	\subsection{Interaction of a spinning black hole with a single stationary infinite string}
	The angular momentum of the black hole ${\bf J}$ evolves due to the torque applied by the string. For the case with a single string entering the horizon,
	we can use Eq.~(\ref{torque}) to write down the evolution equation in vector form
	\begin{equation}
		{d{\bf J}\over dt}=-{\mu \over R}\left[{\bf J}-\left({\bf n}\cdot {\bf J}\right){\bf n}\right],
		\label{evol}
	\end{equation}
	where ${\bf n}$ is the unit vector along the string at $R\ll r\ll L$. This implies
	\begin{eqnarray}
		{\bf J}_\parallel&=&{{\bf J}_\parallel}_0,\\
		{\bf J}_\perp &=& {{\bf J}_\perp}_0\exp\left[-{\mu\over R}t\right],
		\label{jcomps}
	\end{eqnarray}
	where ${\bf J}_\parallel$ and ${\bf J}_\perp$ are components of the angular momentum vector parallel and perpendicular to the string, respectively. Therefore the angular momentum vector of the black hole aligns with the string on a timescale $t_\mu=R/\mu$, while keeping ${\bf J}_\parallel$ fixed. If the black hole is threaded through by a straight string, the alignment would occur on a timescale $t_\mu/2$. We note that while the spin-down of a black hole from a stationary string was noted in, e.g., \cite{2018PhRvD..97b4024B}, the alignment of the spin direction with the string is pointed out here for the first time. 
	
	As the string is stationary, no energy is extracted from the black hole by the string \cite{1972CMaPh..27..283H}. The black hole converts its rotational energy into 
	its  irreducible mass, by increasing its entropy through dissipation at the horizon. The full mass of the black hole stays fixed.
	
	\subsection{Horizon friction}
	
	In the previous subsection we assumed that the tangent unit vector representing the long string at a black hole ${\bf n}$ was stationary. However, since we are exploring a moving string loop, we  also consider the case where ${\bf n}$ is rotating	slowly about the black hole, with the angular velocity ${\bf o}={\bf n}\times\dot{\bf n}$.
	If the black hole is not spinning, this creates torques acting on the black hole, which in the limit $\left|{\bf o}\right|\ll 1/R$ must scale linearly as
	\begin{equation}
		{\bf Q}=\beta{\bf o}.
	\end{equation}
	In order to find the coefficient $\beta$, let's perform a mental experiment in which we add a small spin to the black hole around an axis that is perpendicular to ${\bf n}$. Using Eq.~(\ref{evol}), we can express the total torque acting on the black hole to linear order in ${\bf o}$ and 
	${\bf J}$:
	\begin{equation}
		{\bf Q}=\beta{\bf o}-{\mu\over R}{\bf J}.
		\label{q1}
	\end{equation}
	For $\left|{\bf J}\right|\ll R^2$, the angular velocity of the black hole is given by 
	\begin{equation}
		\mathbf{\Omega}_{\rm BH}={{\bf J}\over 4R^3}.
	\end{equation}
	Therefore,
	\begin{equation}
		{\bf Q}=\beta{\bf o}-4\mu R^2 \mathbf{\Omega_{\rm BH}}.
	\end{equation}
	The torque acting on ${\bf n}$ has to be zero when it is co-rotating with the black hole. Therefore,
	\begin{equation}
		\beta=4\mu R^2.
	\end{equation}
	
	The expression for the torque ${\bf Q}$ allows us to estimate the timescale on which a loop bound to the non-spinning black hole will dissipate and be swallowed by the black hole.
	 The torques $-{\bf Q}_{1,2}$ applied to the string at its ends $1$ and $2$, combined with the motion of those ends, will result in a loss of energy and length by the string:
	\begin{equation}
		{dE\over dt}=-\beta({\bf o}_1^2+{\bf o}_1^2)=-\beta\left[\left({d{\bf n}_1\over dt}\right)^2+\left({d{\bf n}_2\over dt}\right)^2\right].
	\end{equation}
	To order-of-magnitude,
	\begin{eqnarray}
		{dE\over dt}&\sim& -\beta L^{-2}\sim -\mu (R/L)^2,\nonumber\\
		{dL\over dt}&\sim& -(R/L)^2,\label{ldotfriction}\\
		t_{\rm fr}&=&{L\over |dL/dt|}\sim L^3/R^2,\nonumber
	\end{eqnarray}
	where $t_{\rm fr}$ is the timescale for the loop to be absorbed by the non-spinning black hole through horizon friction.
	
	\subsection{Superradiance}
	In this section we show that a circularly polarized tension wave is amplified upon reflection from a spinning black hole, provided its sense of rotation is the same as that of the black hole, and its angular frequency $\omega<\Omega_{\rm BH}\cos\theta$, where $\theta$ is the angle between the string and the spin axis of the black hole. This leads to a very fast growth of short wavelength perturbations on a bound string loop.
	Our treatment gives an exact answer for $\Omega_{\rm BH},\omega\ll 1/R$, but order-of magnitude extrapolation to greater angular frequencies is warranted. A more general treatment is possible but is beyond the scope of our paper. 
	
	Consider a small-amplitude incoming elliptically-polarized wave coming towards the black hole along a string that is straight for $\sigma\gg R$. Mathematically it is given by 
	\begin{equation}
	    {\bf \delta r}_{\rm in}=A_1 {\bf e}_1 \cos[\omega(\sigma+t)]+A_2 {\bf e}_2\sin[\omega(\sigma+t)],
	    \label{incoming}
	\end{equation}
	where ${\bf \delta r}$ is the string displacement from the resting position and ${\bf e}_{1,2}$ are two unit vectors chosen so that (${\bf e}_1,{\bf e}_2, {\bf n_0}$) form a right-handed  orthonormal basis, and that the polarization tensor is diagonalized. Here ${\bf n}_0$ stands for the asymptotic unit vector of the unperturbed string. To zero'th order, the reflected outgoing wave is given by
	\begin{equation}
	    {\bf \delta r}_{\rm out}=-A_1 {\bf e}_1 \cos[\omega(\sigma-t)]+A_2 {\bf e}_2\sin[\omega(\sigma-t)].
	    \label{outgoing}
	\end{equation}
	However, as we presently see, the black hole exchanges energy and angular momentum with the wave, and therefore the amplitudes of the outgoing wave are expected to be slightly different from those of the ingoing wave. 
	
	The incoming fluxes of energy and of the ${\bf n_0}$-component of angular momentum are given by
	\begin{eqnarray}
	    {dE_{\rm in}\over dt}&=&{1\over 2} \mu \omega^2 \left(A_1^2+A_2^2\right)\label{influxE} \\
	    {\bf n}_0\cdot {d{\bf J}_{\rm in}\over dt}&=&\mu \omega A_1 A_2,
	    \label{influxJ}
	\end{eqnarray}
	and similarly for the outgoing wave. We can figure out the incremental changes in the reflected amplitudes by
	computing the time-averaged power and torque applied by the black hole to the wavy string. 
	
	The vector ${\bf n}(t)$ is oscillating periodically, as follows:
	\begin{eqnarray}
	  {\bf n}&=&{\bf n}_0-2\omega\left[ A_1{\bf e}_1\sin(\omega t)- A_2{\bf e_2}\cos(\omega t)\right]\nonumber \\
	    & &-2\omega^2 \left[A_1^2\sin^2(\omega t)+A_2^2\cos^2(\omega t)\right]{\bf n}_0.
	    \label{deltan}
	\end{eqnarray}
	The last term ensures that $|{\bf n}|=1$ up to the second order in $A_1,A_2$. The torque
	applied to the string by the black hole is given by
	\begin{equation}
	    {\bf Q_s}=4\mu R^2\left[\mathbf{\Omega}_{\rm BH}-\left({\bf n}\cdot \mathbf{\Omega}_{\rm BH}\right){\bf n}-{\bf n}\times {d{\bf n}\over dt}\right].
	    \label{Qs1}
	\end{equation}
	The last term on the right-hand side represents horizon friction. When integrating it over
	the wave period, we get twice the directed area of the contour drawn by the ${\bf n}$-vector on the unit sphere:
	\begin{equation}
	    \int_0^{2\pi/\omega}{\bf n}\times {d{\bf n}\over dt} dt=8\pi\omega^2 A_1 A_2{\bf n}_0.
	    \label{horfric1}
	\end{equation}
	Integrating over one cycle, we get
	\begin{eqnarray}
	    \int_0^{2\pi/\omega}{\bf n}_0\cdot {\bf Q}_s dt&=&16\pi\mu R^2\omega\label{intQ}\\
	                 & &\times \left[\Omega_{\rm BH}\cos\theta \mathcal{A}_+
	                  -\omega \mathcal{A}_\times\right],\nonumber 
	\end{eqnarray}
	where 
	\begin{eqnarray}
	    \mathcal{A}_+&=&A_1^2+A_2^2,\label{quad}\\
	    \mathcal{A}_\times&=&2A_1A_2.\nonumber
	\end{eqnarray}
	The time-averaged torque about the string line is given by
	\begin{eqnarray}
	    {\bf n}_0\cdot \langle{\bf Q}_s\rangle&=&8\mu R^2\omega^2\label{Qsav}\\
	    & &\times  \left[\Omega_{\rm BH}\cos\theta \mathcal{A}_+
	                  -\omega \mathcal{A}_\times\right].\nonumber 
	\end{eqnarray}
	Comparing Eqs.~(\ref{Qsav}) and (\ref{influxJ}), we see that upon reflection
	\begin{equation}
	    \Delta \mathcal{A}_\times=16R^2\omega\left[\Omega_{\rm BH}\cos\theta \mathcal{A}_+
	                  -\omega \mathcal{A}_\times\right].
	                  \label{deltacross}
	\end{equation}
	The second relation is obtained from the conservation of energy. The work done by the black hole on the string over one cycle equals
	\begin{equation}
	    \Delta E_s=\int_0^{2\pi/\omega} {\bf Q}_s\cdot\left({\bf n}\times{d{\bf n}\over dt}\right)dt.
	    \label{energyint}
	\end{equation}
	Evaluating this using Eq.~(\ref{Qs1}), we obtain
	\begin{equation}
	    \Delta E_s=16\pi\mu R^2\omega^2\left[\Omega_{\rm BH}\cos\theta \mathcal{A}_\times-\omega\mathcal{A}_+\right].
	\end{equation}
	Comparing this with Eq.~(\ref{influxE}), we get
	\begin{equation}
	    \Delta \mathcal{A}_+=16R^2\omega\left[\Omega_{\rm BH}\cos\theta \mathcal{A}_{\times}-\omega\mathcal{A}_+\right].
	    \label{deltaplus}
	\end{equation}
	It is of interest to consider the eigenmodes of the system, i.e. the waves that do not change their polarization state upon reflection from the black hole. This implies
	$\Delta\mathcal{A}_+/\mathcal{A}_+=\Delta\mathcal{A}_\times/\mathcal{A}_\times$. From Eqs.~(\ref{deltacross}) and (\ref{deltaplus}) we see that this is equivalent to 
	\begin{equation}
	    \mathcal{A}_+=\pm\mathcal{A}_\times,
	\end{equation}
	and 
	\begin{equation}
	    A_1=\pm A_2,
	\end{equation}
	which corresponds to circularly polarized waves. For the string in the northern hemisphere, with $\cos\theta>0$, the $A_1=-A_2$ wave is partially absorbed by the black hole, with
	\begin{equation}
	    {\Delta \mathcal{A}_+\over \mathcal{A}_+}=-16R^2\omega \left[\Omega_{\rm BH}\cos\theta+\omega\right].
	\end{equation}
	On the other hand, the wave with $A_1=A_2$ is amplified if $\omega<\Omega_{\rm BH}\cos\theta$, and is partially absorbed if $\omega>\Omega_{\rm BH}\cos\theta$:
	\begin{equation}
	    {\Delta \mathcal{A}_+\over \mathcal{A}_+}=16R^2\omega \left[\Omega_{\rm BH}\cos\theta-\omega\right].
	    \label{deltaAoverA}
	\end{equation}
	Intuitively this makes sense. If the wave is rotating in the direction opposite that of the BH, horizon friction dominates. The greater  the black hole spin, the stronger is the absorption. If, on the other hand, the wave is rotating in the same direction as the black hole, it is amplified if the black hole spins faster than the wave (measured after projection of $\mathbf{\Omega}_{\rm BH}$ onto the string) and is partially absorbed if the wave spins faster than the black hole. Qualitatively similar results can be derived for an ordinary string viscously interacting with a rotating sphere, in the spirit of the 
	original rotational superradiance proposal by Zeldovich (\cite{1971JETPL..14..180Z}).
	The maximum growth rate is achieved for 
	\begin{equation}
	    \omega={1\over 2}\Omega_{\rm BH}\cos\theta,
	\end{equation}
	    with the relative amplitude increase of
	\begin{eqnarray}
	\left[{\Delta A_{1,2}/ A_{1,2}}\right]_{\rm max}&\simeq&{1\over 2}\left({\Delta\mathcal{A}_+/\mathcal{A}_+}\right)_{\rm max}\label{maxgrowth}\\
	&=&2R^2\Omega_{\rm BH}^2\cos^2\theta. 
	\nonumber
	\end{eqnarray}
	We emphasize again that this result is exact in the limit $\Omega_{\rm BH}R\ll 1$, but 
	we expect that both the criterion for superradiance and the expression for the maximum growth rate are correct to order of magnitude for all values of $\Omega_{\rm BH}$.
	
	\subsection{Black-hole bomb}
	A wave reflected from the black hole along a straight string will never come into contact with the black hole again. This is not so if the string is a bound loop with both ends attached to the black hole. Consider a pulse of circularly polarized waves  with angular frequency $\omega<\Omega_{\rm BH}\cos\theta$ that is reflected from the black hole when the corresponding string end is making a polar angle $\theta$ with the black hole spin axis.
	If the waves in the pulse are rotating in the same sense as the black hole, then the pulse is  amplified upon reflection from the black hole. As the pulse travels to the other end of the string,
	it approaches the black hole along the same direction along which it was traveling upon the first reflection. This is because ${\bf n}_2(t+L)=-{\bf n}_1(t)$. Since the helicity of the wave is conserved as it travels along the loop, the pulse is rotating in the same sense as the black hole upon the second approach, and therefore it is amplified upon the second reflection also. The amplification repeats with each bounce, leading to an exponential growth of the
	pulse with the timescale
	\begin{equation}
	    t_{\rm sr}=\left[8R^2\omega(\Omega_{\rm BH}\cos\theta-\omega)\right]^{-1} L.
	    \label{tsr}
	\end{equation}
	The fastest growth will take place for the wave's angular frequency
	\begin{equation}
	    \omega_0={1\over 2}\Omega_{\rm BH}\cos\theta_{\rm min}
	    \label{omega0}
	\end{equation}
	where $\theta_{\rm min}$ is the smallest angle between the spin axis and the string end.
	The pulse of wavelength $\lambda_0=2\pi/\omega_0=4\pi/(\Omega_{\rm BH}\cos\theta_{\rm min})$ and timed to reflect from the black hole when $\theta=\theta_{\rm min}$, will grow on a timescale
	\begin{equation}
	    t_{\rm bomb}={L\over 2(R\Omega_{\rm BH}\cos\theta_{\rm min})^2}.
	\end{equation}
	The subscript in the equation above refers to the concept of a black-hole bomb, originally conceived by Press \& Teukolsky (\cite{1972Natur.238..211P}). They pointed out that if a spinning black hole was placed in a cavity with reflecting walls, then there would be exponentially amplified superradiant modes. In our case, the string loop itself plays a role of the cavity, causing repeated superradiant interaction between the black hole and pulses of tension waves.
	
	The growth timescale in Eq.~(\ref{tsr}) is likely to be shorter than all other evolutionary timescale for the loop, and therefore one can expect the short-wavelength 
	waves to quickly reach a non-linear amplitude on the string. What happens afterwards is not
 clear from the arguments given above, and a new approach is needed. In the next section we make a step in this direction, by exploring the evolution of the auxiliary curve. We finish this section by a discussion
 of angular momentum exchange between the black hole and the string loop.

	\subsection{Black hole spindown}
	Since the loop has two ends attached to the black hole, the angular momentum equation (\ref{evol}) is modified to
	\begin{equation}
		{d{\bf J}\over dt}=-{\mu \over R}\mathcal{N}{\bf J},
		\label{evol1}
	\end{equation}
	where $\mathcal{N}$ is a linear evolution operator defined by
	\begin{equation}
		\mathcal{N}{\bf J}=2{\bf J}-\left(\bf{n}_1\cdot {\bf J}\right){\bf n}_1-\left(\bf{n}_2\cdot {\bf J}\right){\bf n}_2.
		\label{mathN}
	\end{equation}
	Recall that ${\bf n}_1(t)$ and ${\bf n}_2(t)$ are the unit vectors at the 2 ends of the string that are pointing away from the black hole, which are given by 
	\begin{eqnarray}
		{\bf n}_1(t)&=&{\bf a}^{\prime}(-t)\nonumber\\
		{\bf n}_2(t)&=&-{\bf a}^{\prime}(L-t).
		\label{n1n2}
	\end{eqnarray}
	As was already noted above, ${\bf n}_1$ and $-{\bf n}_2$ trace out the same trajectory during the loop's oscillation period, but with a half-period delay. Thus the oscillation-averaged angular momentum evolution operator is given by 
	\begin{equation}
		\langle\mathcal{N}\rangle {\bf J}=2\left\{{\bf J}-{1\over 2L}\int\limits_0^{2L} \left[{\bf a}^{\prime}(t)\cdot{\bf J}\right]{\bf a}^{\prime}(t) dt \right\}.
	\end{equation}
	If vectors ${\bf n}_1$ and ${\bf n}_2$ are not pointing along the same line, the operator $\mathcal{N}$ is positive-definite with no eigenvalues that are equal $0$. Therefore all three components of angular momentum ${\bf J}$ will reduce exponentially, on a timescale
	\begin{equation}
	    t_{\rm spindown}\sim R/\mu.
	    \label{tspindown}
	\end{equation}
	Note that this timescale is similar to that of $t_{\rm shrink}$ from Eq.~(\ref{tshrink}), which is the timescale for loop depletion due to the black hole's finite mass.  
	
	The lost black hole angular momentum will be acquired by the string loop; its angular momentum ${\mathbf{\Lambda}}$ will evolve according to the equation
	\begin{equation}
		{d{{\mathbf{\Lambda}}}\over dt}={\mu\over R}\langle\mathcal{N}\rangle{\bf J},
		\label{amloop}
	\end{equation}
	where we averaged over the loop's oscillation period.
	Let $z$-axis point in the direction of the black hole spin. It is clear that 
	\begin{equation}
		{d\Lambda_z\over dt}
		\sim {\mu R\alpha}>0,
		\label{amloop1}
	\end{equation}
	where $\alpha=a/R$ is the dimensionless spin of the black hole. Since $\Lambda\sim \mu L^2$, the loop's angular momentum will change on a timescale
	\begin{equation}
		t_{\rm am}\sim {L^2\over \alpha R}.
		\label{tam}
	\end{equation}
	Over time $t>t_{\rm am}$, the component $\Lambda_z$ becomes positive. 
	
	We note that the computation above did not consider the angular momentum drained from the black hole by a superradiant wave. It is easy to check, however, that
	\begin{equation}
	    {(dJ/dt)_{\rm wave}\over (dJ/dt)_{\rm loop}}\sim (A\omega)^2,
	\end{equation}
	and thus the two contributions become comparable only when $A\omega\sim 1$, i.e. in a strongly non-linear regime. Similarly, the change of rotational energy of the black hole due to a superradiant wave is related to the total loss of rotational energy by
	\begin{equation}
	    {(dE_{\rm rot}/dt)_{\rm wave}\over (dE_{\rm rot}/dt)_{\rm total}}\sim (A\omega)^2
	\end{equation}
	for $A\omega\lesssim 1$.
	Therefore, the fraction of the black hole's rotational energy that gets converted into
	string is determined by the nonlinear saturation of the string superradiance. The rest of the rotational energy is converted into the irreducible mass of the black hole\footnote{It is easy to see that the most efficient, albeit the slowest spin  energy conversion into string length takes place if the string nearly co-rotates with the black hole. In that case the black hole is spun down nearly adiabatically, with vanishing increase in its area/entropy.}. 
	
	To make further progress, we need to consider the non-linear evolution of the string loop due to its 
	interaction with a spinning black hole.

\section{Evolution of the auxiliary curve}
 In section 2 we saw that the trajectory of a loop pinned to a point is described by a a stationary fixed auxiliary curve ${\bf a}(\sigma)$, with $0\le\sigma\le 2L$ and ${\bf a}(0)={\bf a}(2L)$. If instead a loop is anchored on a spinning black hole, its periodic orbit is changing slowly on a timescale much longer than a single oscillation period $2L$. It is attractive to think of this in terms of slow deformation of the auxiliary contour ${\bf a}$. Suppose the contour is deforming with velocity ${\bf v}(\sigma)$, where it makes sense to restrict ${\bf v}(\sigma)$ to be perpendicular to the curve's tangent ${\bf a}^{\prime}(\sigma)$. Recalling Eq.~(\ref{EJloop}), we can write down the rate of change of the
 loop's angular momentum:
 \begin{equation}
     {d{\bf \Lambda}\over dt}=-{1\over 2}\mu \oint_0^{2L} {\bf v}\times {\bf a}^{\prime}d\sigma.
     \label{torqueloop}
 \end{equation}
 This needs to be compared to the expression in Eq.~(\ref{Qs1}) for the torque applied to the string. Integrating this over an oscillation period, recalling ${\bf n}(t)={\bf a}^{\prime}(-t)$, and multiplying by $2$ to account for the two ends of the loop attached to the black hole, we obtain the change of the strings angular momentum over an oscillation period:
 \begin{equation}
     \Delta{\bf \Lambda}=-8\mu R^2\oint_0^{2L} \left[{\bf a}^{\prime}\times\mathbf{\Omega}_{\rm BH}+{\bf a}''\right]\times {\bf a}' d\sigma.
 \end{equation}
 In order to make the two expression consistent, one needs to choose
 \begin{equation}
     {\bf v}(\sigma)={8 R^2\over L}\left[{\bf a}^{\prime}(\sigma)\times\mathbf{\Omega}_{\rm BH}+{\bf a}''(\sigma)\right],
     \label{geomflow}
 \end{equation}
 where the derivatives are understood to be evaluated with respect to the length along the curve. Clearly, as the curve evolves under the action of this flow, $L$ changes as well. 
 
 The change of length in the auxiliary curve is given by
 \begin{equation}
     {dL_a\over dt}=-\oint_0^{2L} {\bf v}\cdot{\bf a}''d\sigma.
 \end{equation}
 One can check that with the expression for ${\bf v}$ above, one  obtains the average rate of energy
 change $dE/dt=0.5\mu~(dL_a/dt)$ that is consistent with what one would obtain from Eq.~(\ref{energyint}) where the integration is over an oscillation period $2L$.
 
 \subsection{Curve-shortening flow.} 
 Equation (\ref{geomflow}) describes the non-linear evolution of the auxiliary curve, with the non-linearity implicit due to $\sigma$ being understood as the length coordinate. It represents a one-dimensional geometric flow, and as such is of considerable interest to mathematicians. In fact, when $\Omega_{\rm BH}=0$,
 the equation of motion
 \begin{equation}
     {\bf v}\propto{\bf a}''
 \end{equation}
 describes a famous and extensively studied {\it curve-shortening flow} (see, e.g., \cite{1984InMat..76..357G},\cite{gage1986},\cite{10.2307/1971486}). For practical purposes, the main result of this exploration is something of which one can immediately convince oneself through qualitative arguments and simple numerical experiments (we have done both): asymptotically, the curve becomes a shrinking planar circle and disappears in a singularity after a finite time. A circular auxiliary curve of length $2L$ corresponds to a degenerate physical loop that extends radially from the black hole to radius $L/\pi$ and then traces the same radial line back to the black hole. This {\it double-line} is rotating around the 
 black hole with the angular velocity $\pi/L$; the tip of the double-line is moving with the speed of light
 and thus the double-line is a prolific emitter of gravitational waves. The length of the line is shrinking due to horizon friction and the line disappears after the time 
 \begin{equation}
     t_{\rm short}\simeq {L^3\over 24\pi^2 R^2},
     \label{tshort}
 \end{equation}
 cf.~Eq.~(\ref{ldotfriction}). In the process the double-line makes
 \begin{equation}
     N_{\rm short}\simeq {1\over 2}\left({L\over 4\pi R}\right)^2
 \end{equation}
 turns around the black hole.
 The approximate equality in the equation above is due to the fact that our curve-shortening formalism is only valid for $L\gg R$. 
 
 \subsection{Curve-lengthening due to black-hole spin.}
 Consider now a non-self-intersecting auxiliary curve in the  plane perpendicular to $\mathbf{\Omega}_{\rm BH}$, with the circulation in the same sense as rotation of the black hole. Clearly the effect of the first
 term on the right in Eq.~(\ref{geomflow}) is to expand the curve outwards and make it increasingly more 
 round. If only the first term is included in the evolution equation, some of the initial configurations 
 develop kinks and singularities, however the inclusion of the second term smooths them out.
 
 If one can neglect the second term (a good approximation when ${\Omega}_{\rm BH}\gg 1/L$), then one
 can prove the following. Suppose the auxiliary curve is convex and everywhere differentiable. Choose the point
 of origin inside the curve and find maximum and minimum radii $a_{\rm max}$ and $a_{\rm min}$. As the curve lengthens, the diffarence $a_{\rm max}-a_{\rm min}$ remains constant. Therefore the ellipticity 
 \begin{equation}
     \epsilon\equiv{a_{\rm max}-a_{\rm min}\over a_{\rm max}+a_{\rm min}}\sim 1/L,
 \end{equation}
 and therefore the loop becomes increasingly circular. The latter expands according to
 \begin{equation}
     L=\sqrt{L_0^2+16\pi R^2\Omega_{\rm BH} t},
     \label{LT}
 \end{equation}
 where $L_0$ is the initial invariant length of the loop with circular auxiliary curve.
 
 A  circular auxiliary curve corresponds to a double-line which in this case is rotating around the black hole in its equatorial plane and is growing in length, while reducing its angular velocity of rotation.
 If the double-line is rotating at the same angular velocity as the black hole (and thus has $L=\pi/\Omega_{\rm BH}$), there is no energy and angular momentum exchange with the black hole and $L$ does not change. This equilibrium, however, is unstable: double lines longer/shorter than $\pi/\Omega_{\rm BH}$ will lengthen/shorten in their extent.
 
 \subsection{Superradiance revisited.}
 It is  straightforward to compute the growth rate for superradiant modes by considering a helical short-wavelength perturbation on a part of the auxiliary curve that can be considered locally straight. For simplicity, let's assume that the straight part is along the spin axis, and choose $z$-axis to be also aligned with the spin. The helical wave can be written as 
 \begin{eqnarray}
     {\bf a}_x(\sigma)&=&A\cos\left({\omega\over \sqrt{1+A^2\omega^2}}\sigma\right),\nonumber\\
     {\bf a}_y(\sigma)&=&A\sin\left({\omega\over \sqrt{1+A^2\omega^2}}\sigma\right),\label{hwave}\\
     {\bf a}_z(\sigma)&=&{1\over \sqrt{1+A^2\omega^2}}\sigma,\nonumber 
 \end{eqnarray}
 where it is understood that the above expression is valid only in some small part of the auxiliary loop (please note that ${|\bf a}'|=1$). The amplitude of the helical perturbation evolves according to the following equation:
 \begin{equation}
     {dA\over dt}={8R^2\omega A\over L}\left[{\Omega_{\rm BH}\over \sqrt{1+A^2\omega^2}}-{\omega\over 1+A^2\omega^2}\right].
 \end{equation}
 In the linear case $A\omega\ll 1$, the amplitude grows exponentially with the timescale
 \begin{equation}
     t_{\rm sr}=\left[8R^2\omega(\Omega_{\rm BH}-\omega)\right]^{-1}L,
 \end{equation}
 which is identical to Eq.~(\ref{tsr}) for the case $\theta=0$.
 In the non-linear case $A\omega\gg 1$, the auxiliary curve is tightly winding up the $z$-axis with its tangent nearly horizontal. For $\Omega_{\rm BH}\gg 1/A$, the amplitude grows at a constant rate so long as the overall-length of the curve has not changed much,
 \begin{equation}
     {dA\over dt}={8R^2\Omega_{\rm BH}\over L},
 \end{equation}
 with remarkable independence from  $\omega$.
 
 The general lesson from the above discussion is that the curve-lengthening due to black-hole spin will amplify 
 any wiggle on the curve that has the right helicity into a nearly horizontal and nearly closed circular arc.
 Because of superradiance, we expect $\alpha L/R$ of such nearly-circular segments to develop. 
 
\subsection{Self-intersection of the auxiliary curve and production of new bound loops.}
The curve-lengthening described in previous paragraphs will likely lead to multiple (of order $\alpha L/R$) self-intersections of the evolving auxiliary curve, provided that the physical reconnections of the loop will
not drastically alter the picture. Suppose that loop reconnections do not take place, as could be the case if the loop is made of a super-string. What then is the meaning of self-intersections of the auxiliary curve?

Suppose ${\bf a}(\sigma_1)={\bf a}(\sigma_2)$. Then for $t=-(\sigma_1+\sigma_2)/2$ and $\sigma=|\sigma_2-\sigma_1|/2$, ${\bf r}(\sigma,t)=0$. This is not surprising, since  ${\bf r}(\sigma,t)$ span all possible directed half-chords of the auxiliary curve, which must include $0$ if the curve is self-intersecting. Physically, this means that some middle part of the loop gets captured by the black hole and the loop splits into two loops both attached to the black hole.

One may wonder whether the capture actually happens, since the self-intersection of the auxiliary curve is instantaneous while it takes 
time $L$ for the loop to complete a half-osciallation during which the capture would take place\footnote{To make this concern more precise, consider the moment that auxiliary curve self-intersects. If at that moment the evolution of the auxiliary curve is switched off, from an argument above we see that a portion
of the loop would be captured within the half-cycle from that moment. However, if the curve keeps evolving, then it is not a-priori clear that it would not evolve far enough during the
half-cycle, so that the string would just miss the black hole.} . We note however, that a black hole has a Schwarzschild radius $R_s=2R$. From Eq.~(\ref{geomflow}) we see that the maximal speed with which the curve moves is $2R_s^2\Omega_{\rm BH}/L$; during half-oscillation time it moves by a distance no greater than $2R_s$ (recall $\Omega<1/R_s$). Since the radius is a half-chord, even if the two segments of the curve cross with maximal possible velocity, there will be a half-oscillation interval such that
the distance of the closest approach of the loop to the black hole is less than $2R_s$. As numerical experiments of \cite{1999CQGra..16.2403D} show, strings with such small impact parameter with respect to a black hole typically get captured. Thus the string capture is overwhelmingly likely, especially for non-maximally-spinning black holes.
Thus remarkably, without string reconnections the initially captured loop will split into into $\sim \alpha L/R$ loops independently bound to the spinning black hole. Each of the bound loops will evolve into a double-line rotating close to the equatorial plane.

\subsection{Asymptotically expanding double lines and reconnections.} 

At a first glance, expanding double lines that correspond to nearly-circular auxiliary curves, do not appear stable if any amount of reconnection is present in the system. However, we argue that this intuition could be misleading. Below, 	we obtain  a general asymptotic form for a nearly-circular
auxiliary curve that is expanding due to the spin-induced curve lengthening. We show that it is straightforward to find examples of such solutions that never self-intersect.
	
\paragraph{Asymptotic loop.}

By appropriate rescaling of time and spatial dimension in Eq.~(\ref{geomflow}), we obtain the dimensionless evolution equation for the auxiliary curve, for a BH rotating around the z-axis: 
\beq
\partial _t{\bf a}=\partial _\sigma^2{\bf a}-\hat{z}\times\partial _\sigma{\bf a}+u\partial _s{\bf a},
\eeq
where $d\sigma=|\partial _s{\bf a}|ds$ is the increment of the auxiliary-curve length at a fixed time $t$, and $s$ is a parameterization of the curve.  Here we assume $0<s<2\pi$, so that ${\bf a}(t,s)$ is a $2\pi$-periodic function of $s$. The last term on the right-hand side contains an arbitrary $2\pi$-periodic function of $s$, $u(t,s)$, which accounts for a continuous arbitrary re-parameterization of the curve. 

For a long auxiliary curve, the first term in the right-hand side can be dropped. Since we expect an asymptotic solution that is close to circular, we parameterize the curve in cylindrical coordinates ${\bf a}(s)=[\rho(s),\phi(s),z(s)]$ by choosing $s=\phi$. 
Denoting $\partial _t$ and $\partial _\phi$ by the ``dot'' and ``prime'', we get
\beq
(\dot{\rho},0,\dot{z})=-\frac{(0,0,1)\times(\rho',\rho,z')}{\sqrt{\rho'^2+\rho^2+z'^2}}+(\rho',\rho,z')u.
\eeq
The second of these three equations gives $u$, and then the other two equations read
\beq
\dot{\rho}=\frac{\rho'^2+\rho^2}{\rho\sqrt{\rho'^2+\rho^2+z'^2}}
\eeq
\beq
\dot{z}=\frac{\rho'z'}{\rho\sqrt{\rho'^2+\rho^2+z'^2}}
\eeq

An expanding circle is an exact solution, $\rho=t_0$, $z=0$, where $t_0 $ is the ``age'' in terms of the rescaled time and more usefully,   the radius of the auxiliary curve. Therefore, the evolution of a slightly deformed circle, to the first two non-vanishing orders, is given by $\dot{\rho}=1$, $\dot{z}=0$, and we get 
\beq
\rho=t_0+h(\phi),~~~z=g(\phi).
\eeq
Since we are only interested in the shape of the curve, we can rescale the axes by $t_0$. The late-time asymptotic form of the auxiliary curve can be written as
\beq\label{asac0}
\rho=1+h(\phi)/t_0,~~~z=g(\phi)/t_0.
\eeq
 Equation (\ref{asac0}) is valid up to the first order in $1/t_0$. In terms of the length parameter $\sigma$, again up to the first order in $1/t_0$, with ${h}(\phi)\equiv f'(\phi)$, we get
\beq\label{asac}
\rho=1-f'(\sigma)/t_0,~~~\phi=\sigma+f(\sigma)/t_0,~~~z={g}(\sigma)/t_0.
\eeq
This is the general late-time form of the  nearly-circular expanding
auxiliary curve rescaled by its radius.

\paragraph{Non-self-intersecting asymptotic loop.}

Recall that the string loop ${\bf r}(t,\sigma)$ can be expressed through the auxiliary curve ${\bf a}(\sigma)$:
\beq
2{\bf r}(-t,\sigma)={\bf a}(t+\sigma)-{\bf a}(t-\sigma).
\eeq

It is convenient to use Cartesian coordinatization $(a_x,a_y,a_z)$ of the curve. From Eq.(\ref{asac}) (to first order in $1/t_0$) we have
 \beq
a_x+ia_y=\left(1-{f'\over t_0}+i{f\over t_0}\right)e^{i\sigma}.
\eeq
Then for the string loop, in Cartesian $(x,y,z)$, we have
\begin{eqnarray}\label{slxy}
2(x+iy)&=&e^{it}\left\{2i\sin\sigma\right.\nonumber\\
    & &\left.+e^{i\sigma}\left[if\left(t+\sigma\right)-f'(t+\sigma)\right]/t_0\right.\\
     & &\left.-e^{-i\sigma}\left[if(t-\sigma)-f'(t-\sigma)\right]/t_0\right\}.\nonumber
\end{eqnarray}
If the string loop self-intersects at time $t$ at points $\sigma$ and $\tilde{\sigma}$, then to zeroth order in $1/t_0$
\beq
\sin \sigma=\sin \tilde{\sigma} ~~~\Rightarrow ~~~ \tilde{\sigma}=\pi-\sigma,
\eeq
and this value can be used when calculating the first-order terms of the string loop. In the zeroth order term of the string loop, $2i\sin\tilde{\sigma}$, the first order correction of $\tilde{\sigma}$ will make the imaginary parts of the bracket in Eq.(\ref{slxy}) match at $\tilde{\sigma}$ and at $\sigma$ up to first order. Only the real part matching leads to an equation for $f$:
\beq
F(t,\sigma)=F(t,\pi-\sigma),
\eeq
\begin{eqnarray}
F(t,\sigma)&\equiv& \cos\sigma\left[ f'(t+\sigma)-f'(t-\sigma) \right]\nonumber\\
           & &+\sin\sigma \left[f(t+\sigma)+f(t-\sigma)\right],
\end{eqnarray}
or
\begin{align}\label{slin}
\cos\sigma\left[f'(t+\sigma)-f'(t+\sigma-\pi)\right.\nonumber\\
\left.-f'(t-\sigma)+f'(t-\sigma+\pi)\right]+\\
\sin\sigma \left[f(t+\sigma)-f(t+\sigma-\pi)\right.\nonumber\\
\left.+f(t-\sigma)-f(t-\sigma+\pi)\right]=0\nonumber
\end{align}
Recalling that $f(\sigma)$ is a $2\pi$-periodic function of $\sigma$, we see that Eq.(\ref{slin}) is an identity for all even-m Fourier harmonics, $\cos(m\sigma)$ and $\sin(m\sigma)$. Assuming that $f$ has only odd-m harmonics, we can simplify Eq.(\ref{slin}):
\begin{eqnarray}\label{slin1}
\cos\sigma\left[f'(t+\sigma)-f'(t-\sigma)\right]&+&\nonumber\\
\sin\sigma \left(f(t+\sigma)+f(t-\sigma)\right)&=&0.
\end{eqnarray}
For a self-intersection the string loop z-components must also match, $z(t,\sigma)=z(t,\pi-\sigma)$. We will assume that $g$ has only the even-m harmonics (as the odd-m ones don't contribute), and get the second necessary condition for the self-intersection
\beq\label{slin2}
g(t+\sigma)-g(t-\sigma)=0.
\eeq

Now we can give the simplest possible example of a non-self-intersecting string loop. We take $f(\sigma)=\cos(3\sigma)$, because Eq.(\ref{slin1}) is an identity for $m=1$.  We take $g(\sigma)=\sin(2\sigma)$. Then Eqs.(\ref{slin1}, \ref{slin2}) read
\beq
\cos(3t)\cos^3\sigma\sin\sigma=0,~~~\cos(2t)\cos\sigma\sin\sigma=0.
\eeq
This system of equations does not have any physically relevant solutions: $\cos\sigma =0$ is not an option because then $\sigma=\frac{\pi}{2}$ coincides with $\pi-\sigma$; $\sin\sigma =0$ is not an option because $\sigma=0$ and  $\sigma=\pi$ are the points where the string loop attaches to the black hole; the only remaining option is 
\beq
\cos(3t)=0,~~~\cos(2t)=0.
\eeq
which is impossible. Numerically we have confirmed that this asymptotic loop indeed does not self-intersect.

The existence of non-intersecting asymptotic solution means that there exist loop configurations
that will inflate until the black hole is spun down, with $L\sim R\mu^{-1/2}$. Whether such non-intersecting asymptotic states can be reached generically through curve-lengthening combined with reconnections is an open question.

\section{Evolution of a bound loop}
In this section we explore the implications of the results from previous sections on the evolution of the loop bound to the black hole.
We distinguish two cases, that of a non-rotating and rotating black hole. In the former case, our estimates and conclusions are reliable (if a bit boring for astrophysics), while in the
latter case our conclusions are tentative since we do not yet have a detailed understanding of the role of reconnections on the evolution of the loop that is being inflated by a spinning black hole.

\subsection{Non-rotating black hole}
Three processes lead to depletion of the bound loop:

1. Change of the loop's orbit and subsequent ejection, due to the finite mass of the black hole. If the reconnection probability $p>\mu L/R$, 
the loop length decreases approximately exponentially [see Eq.~({\ref{tshrink})], on a timescale
\begin{equation}
    t_{\rm shrink}\sim R/\mu.
\end{equation}
If $p<\mu L/R$, the loop shrinks on a timescale $\sim L/p$.

2. Horizon friction depletes the loop on a timescale 
\begin{equation}
    t_{\rm fr}\sim 5\times 10^{-3} L^3/R^2,
\end{equation}
cf.~Eq.~(\ref{tshort}).
The auxiliary curve shortening caused by the friction will drive the loop towards a double-line
configuration on the same timescale. Since the tip of an ideal double-line moves with the speed
of light, one expects copious production of gravitational waves.

3. Gravitational radiation depletes the loop on a timescale
\begin{equation}
    t_{\rm GW}\sim {L\over \Gamma \mu \mathcal{M}},
\end{equation}
where $\Gamma\sim 50$ is the numerical factor computed for a typical loop with finite number of cusps, and $\mathcal{M}\sim 1$ for a typical loop and is logarithmically large for the double line due to the ultrarelativistic motion of its tip \cite{1997PhRvD..55.6054M}.

For simplicity, let's assume reconnections are efficient.
For $L>(5\times 10^{-3}\mu)^{-1/3}R$, process $1$ dominates over process $2$, while for $L<(5\times10^{-3}\mu)^{-1/3}R$
process $2$ dominates over process $1$. For
\begin{equation}
    \mu<200(\Gamma\mathcal{M})^{-3}\sim 2\times 10^{-3}\mathcal{M}^{-3},
\end{equation}
the gravitational radiation is not the dominant mechanism for the loop depletion for any $L$, with respect to the combination of two other processes. For $\mathcal{M}\lesssim 100$, this criterion is satisfied extremely well for observationally allowed values of $\mu$.

Therefore a loop with the initial size $L_0\gg (5\times10^{-3}\mu)^{-1/3}R$ evolves in two stages. First, it shrinks by  ejection of the daughter loops due to the finite mass of the black hole,
\begin{equation}
    \log L\sim \log L_0- \chi {\mu\over R} t,
    \label{timeevol1}
\end{equation}
where $\chi\sim 1$. After reaching $L\sim (5\times 10^{-3}\mu)^{-1/3}R$, the evolution proceeds by horizon friction and the loop turns into a double line before being swallowed by the black hole. If $t_{\rm death}$ is the time at which the loop disappears, then before that the loop evolves according to
\begin{equation}
    L\sim R^{2/3} (t_{\rm death}-t)^{1/3}.
\end{equation}
In this stage of evolution, the energy of the loop is absorbed into the irreduceable mass of the black hole.

\subsection{Rotating black hole}
We have much less certainty in assessing the loop's evolution if the black hole is rotating, because while we understand the spin-driven lengthening of the auxiliary curve, we do not 
know how reconnections would affect this evolution.
One reasonable guess is that the superradiance of short-wavelength helical tension waves will cause reconnection at a distance $\lesssim 1/\Omega_{\rm BH}$ from the black hole, with the remaining bound loop dissipating due to the horizon friction.  For this size, the auxiliary curve shortening due to horizon friction occurs on a timescale comparable to that of the curve lengthening due to the black hole spin. One can then suppose that in some cases the shrinking wins and the loop gets swallowed by the black hole, and in other cases the
curve lengthening wins and the loop grows by extracting the rotational energy of the black hole.
What exactly reconnections do to the expanding loop needs to be explored. While asymptotic
non-intersecting solutions do exist, it is far from certain that they will be reached through
reconnections. One can imagine limit cycles, where the loop grows, reconnects, ejects a sub-loop,
grows again, etc. As our code is not powerful enough to efficiently find reconnections of loops with changing shapes, this will have to be explored in future work.

If a loop reaches a non-self-intersecting asymptotic form, its invariant length will grow as $\propto \sqrt{t}$, according to Eq.~(\ref{LT}). The black-hole spindown will limit the maximum to which the loop's spacial extent can grow:
\begin{equation}
    L_{\rm max}\sim \left({\alpha_0/\mu}\right)^{1/2}R.
\end{equation}
where $\alpha_0$ is the initial dimensionless spin of the black hole. At $L\gg 1/\Omega_{\rm BH}$, the conversion  of the black hole spin energy into the loop length becomes inefficient. Only a fraction
$\sim 2\pi (L\Omega_{\rm BH})^{-1}$ of the spin energy gets converted into the loop length, and the rest adds to the irreduceable mass of the black hole. Eventually the black hole  becomes virtually non-rotating. This could take place in the real Universe if 
\begin{equation}
    \mu\gtrsim 4\times 10^{-16}~{10^{17}\hbox{s}\over t_0}~{M\over M_{\rm SgrA*}},
\end{equation}
where $t_0$ is the age of the black hole, and $M_{\rm SgrA*}\simeq 4\times 10^6M_\odot$ is the mass of the supermassive black hole at the center of our Galaxy.

This picture is different if the black hole is accreting from a thin disc that is spinning it up.
For
\begin{equation}
    \mu\lesssim \dot{M}\sim 5\times 10^{-15}~{M\over M_{\rm SgrA*}}~{\dot{M}\over \dot{M}_{\rm Ed}},
    \label{mueq}
\end{equation}
the spin-down torque from the string is unable to compensate the spin-up accretion torque. Here $\dot{M}$ is the accretion rate and $\dot{M}_{\rm Ed}\sim 10^{-8} M_\odot\hbox{yr}^{-1}~(M/M_\odot)$ is the Eddington accretion rate. Please note that in geometric units, $\dot{M}$ is a dimensionless quantity. If $\mu\gtrsim \dot{M}$, the dimensionless spin of the black hole saturates at
\begin{equation}
    \alpha_{\rm eq}\sim \dot{M}/\mu
        \label{alphaeq}
\end{equation}
In this case a  fraction of the rest mass of the accreted material is converted into string length and the loop grows indefinitely.

For sufficiently small values of the reconnection probability $p$, the auxiliary curve will develop self-intersections and the loop will break
into smaller loops, all bound to the black hole. If all the smaller loops survive reconnections and
settle into asymptotic non-self-intersecting double lines, the latter will rotate around the
black hole in near-equatorial planes, each with its own angular velocity.
An important effect would be an enhanced slow-down of the black hole, by a factor that equals to the number of the attached double-lines.


While the physics of black holes' interaction with string loops is interesting,  are they likely to ever meet in the real world if strings do exist? In the next section we argue in the affirmative, for reasonable parameters of the string network.

\section{Cosmological considerations}

The probability for a black hole to have a string attached to it crucially depends on whether the black holes are primordial or they are formed by gravitational collapse and accretion in the late universe.  We will mostly focus on the latter possibility and will only briefly comment on primordial black holes in section VII.C.

\subsection{String evolution and capture}

Numerical simulations of cosmic string evolution indicate that strings evolve in a self-similar manner. A Hubble-size volume at any time $t$ contains a few long strings stretching across the volume and a large number of closed loops of size $L \ll t$ (for an up to date review of string simulations, see \cite{2020PhRvD.101j3018B} and references therein).
Long strings move, typically at mildly relativistic speeds ($v\sim 0.2$), and reconnect when they cross.  Reconnections lead to the formation of closed loops.  The loops oscillate periodically and emit gravitational radiation at the rate
\begin{equation}
{\dot E}= \Gamma \mu^2,
\end{equation}
where $\Gamma\sim 50$ is a numerical factor depending on a particular loop configuration.  As loops loose their energy, they gradually shrink and eventually disappear.  The lifetime of a loop of invariant length $L$ is $\tau\sim{L}/{\Gamma \mu}.$
The smallest and most numerous loops surviving at cosmic time $t$ have length
\begin{equation}
L_{\rm min}(t)\sim \Gamma \mu t.
\end{equation}  
Such loops, which are near the end of their lives, acquire 
large velocities, typically $v\sim 0.1$, due to the asymmetric emission of gravitational waves  (this is the so-called gravitational rocket effect.)  These velocities are too high for the loops to be bound to galaxies.  However, larger loops move slower and, if they are longer than a certain length, they can be captured in galactic halos during the epoch of galaxy formation.  Such galactic loops are also the best candidates for capture by a black hole.

Loop clustering in galaxies was originally studied by Chernoff in \cite{2009arXiv0908.4077C}. In that work however, the gravitational rocket effect on the loops before their capture into the halos was not properly taken into account. Recently \cite{2020arXiv200615358J} performed an analysis where the rocket effect was fully accounted for, and showed that the original computations overestimated the number of captured loops by orders of magnitude. Still, as we show below, those loops that do get captured can  collide with the black holes inside the halos, at a rate that is astrophysically significant.  According to the calculations in \cite{2020arXiv200615358J}, the smallest and most numerous galactic loops have length
\begin{equation}
L_G\sim 30\Gamma \mu t_0 \sim 5 \times 10^{-8}\mu_{-20}~{\rm pc},
\end {equation}
where $t_0$ is the present cosmic time and $\mu_{-20}\equiv \mu/10^{-20}$.  The number of such loops in the halo of a typical galaxy like the Milky Way is
\begin{equation}
\mathcal{N}\sim 10^{12}\mu_{-20}^{-3/2}~\eta(p).
\end{equation}
Here the function $\eta(p)$ reflects the fact that low-$p$ networks are more efficient in producing sub-horizon loops; $\eta(1)=1$. Numerical simulations in  \cite{2006PhRvD..73d1301A}  give $\eta(p)\propto p^{-\zeta}$, where $\zeta\sim 0.6$ with considerable uncertainty.

Let's estimate the rate of capture of loops by a supermassive black hole of the 
mass
\begin{equation}
    M=4\times 10^6M_\odot~{M\over M_{\rm SgrA*}}.
\end{equation}
It is convenient  to  restore $G$ and $c$ for our computations in the rest of this subsection, since geometric units are not suitable for classical galactic dynamics calculations that follow.
What is the distribution of loops near supermassive black holes? A classic argument was given by Young  in \cite{1980ApJ...242.1232Y}, in the context of computing stellar distribution near an adiabatically growing black hole. One can show [see Young's equation (29)] that in a spherically symmetric system 
the steady-state distribution function $f(E,J)$ remains conserved as the potential evolves and thus the energy $E$ changes due to the black hole growth; the angular momentum $J$ is conserved. Here it
must be understood that $E$, $J$, and $f$ are functions of position and momentum.
The distribution function away from the black hole is simply $f\sim v_G^{-3}\mathcal{N/V}$,
where $v_G\sim 200$km/sec is the virial  velocity of the halo and  $\mathcal{V}\sim10^{15}\hbox{pc}^3$ is the halo volume.
The distribution function near the black hole is given by $f\sim {n}_{\rm loops}(r) r^{3/2} (GM)^{-3/2}$; here $n_{\rm loops}$ is the number density of loops. Equating the two we get
\begin{equation}
    n_{\rm loops}(r)={(GM)^{3/2}\over v_G^3 r^{1.5}}\mathcal{N/V}.
\end{equation}
Of  interest for us is the case where $L_G>R$, in which case we need to consider the loops whose centers are within $r\sim L_G$ from the BH. Not all of
them will be captured;  their capture probability in one dynamical time $\sim L_G^{3/2}/(GM)^{1/2}$ is $\sim (R/L_G)(c/v)\sim \sqrt{R/L_G}$, where
$v\sim \sqrt{GM/L_G}$.

The rate of loop captures is given by 
\begin{eqnarray}
    {\rm CR}&\sim& n_{\rm loops}(L_G)\left({GM\over L_G}\right)^{1/2}L_G^2\sqrt{R\over L_g}\nonumber\\
    &\sim&{\mathcal{N}\over \mathcal{V}}{c^4\over v_g^3}R^{5/2}L_G^{-1/2}\label{CRyoung}\\
    &\sim& {\mu_{-18}^{-2}\over 3\times 10^{9}\hbox{yr}}\left({M\over M_{\rm SgrA*}}\right)^{2.5}~[\eta(p)].\nonumber
\end{eqnarray}

The estimate above implies that there are reasonable values of $\mu$ for which all supermassive black holes, including the one in our galactic center, will acquire a loop during their lifetime. Even within the simple model we used, the estimate is uncertain and we want to
point out two caveats. Firstly, even before the black hole formation, the loops are likely to cluster towards the inner halo which would enhance their phase space density near the black hole. Secondly, a merger with another black hole could cause the loops to be ejected from
the nucleus of the galaxy by a slingshot mechanism. We shall for now ignore these complications.

The steep $M$-dependence in the equation above implies that masses of the black holes that do capture strings, could be clustered towards the high end of their range. The details will clearly depend on the mass function of the supermassive black holes
\begin{equation}
    \phi(M)= {dN_{\rm BH}\over dV~d\log M},
\end{equation}
where $N_{\rm BH}$ is the number of black holes, $V$ is volume, and $\log M$ is the decimal logarithm of the black hole mass.
Useful illustrative plots of $\phi(M)$ can be found in, e.g., Figure 7a of \cite{2009ApJ...690...20S} (the grey region corresponds to the  mass function measured at redshift $0$) and Figure 2 of \cite{2012AdAst2012E...7K}. As can be seen from the latter figure, $\phi(M)$ changes slowly from $\sim 0.01\hbox{Mpc}^{-3}$ to $\sim 0.001\hbox{Mpc}^{-3}$ between $10^6M_\odot$ and $10^{8.4}M_\odot$ and falls off more steeply at higher masses, very roughly as $\propto M^{-2}$ between $10^{8.4}M_\odot$ and $10^{9.5}M_\odot$. For a given tension
$\mu_{-18}$, the greatest number of loops will be captured by black holes with masses
\begin{equation}
    M_\mu\sim M_{\rm SgrA*}~\mu_{-18}^{4/5}\eta^{-2/5},
\end{equation}
such that the ${\rm CR}\sim 1/(3\times 10^9\hbox{yr}) $ and the capture probability for
the black hole is of order unity (we assume here that there is no benefit for a black hole to capture $>1$ loop, since after reconnections only one loop will remain bound to the black hole).

\subsection{Gravitational wave signature}
At this stage we are unable to make definitive predictions about gravitational waves, since 
we are unable to model the evolution of the captured loop with reconnections. 
Some general quantitative remarks can however be made.

Firstly, we note that in an ultra-optimistic scenario, all of the spin energy of supermassive black holes can be converted into string loops which in turn convert their energy into gravitational waves. The average mass density of supermassive black holes is $\sim 2\times 10^5M_\odot/\hbox{Mpc}^3$ \cite{2012AdAst2012E...7K}. One can then easily estimate the ratio between the energy density in black holes' spin and the  energy density of the universe:
\begin{equation}
    \Omega_{\rm spin}\sim 10^{-7}\bar{\alpha}^2,
    \label{omegaspin}
\end{equation}
where $\bar{\alpha}$ is the average dimensionless spin of the black holes.
Measurements of x-ray and radio emission from accreting supermassive black holes in galactic nuclei indicate that they are rapidly rotating (e.g., \cite{2020arXiv200808588J}, \cite{2019ApJ...886...37D}, and references therein\footnote{However, recently \cite{2020arXiv200811734F} provided an interesting argument for the upper bound $\alpha\lesssim 0.1$ on the spin of the SgrA* black hole. It was based on the fact that the S-stars near SgrA* appear to belong to two mutually inclined discs \cite{2020ApJ...896..100A}, while the large spin of the black hole would destroy such structures through the Lense-Thirring precession of the orbits \cite{2003ApJ...590L..33L}.}). This is consistent with an argument that the supermassive black holes acquire most of their mass from thin accretion discs \cite{1982MNRAS.200..115S}, \cite{2002MNRAS.335..965Y}, in which case one might expect that a significant fraction of them  rotate rapidly, with $\alpha\sim 1$. 

The estimate in Eq.~(\ref{omegaspin})  is six orders of magnitude above the projected sensitivity of LISA of $\Omega_{\rm GW}\sim 10^{-13}/h^2$ and many orders of magnitude greater than the projected $\Omega_{\rm GW}$ produced by current cosmological models with cosmic strings \cite{2018PhLB..778..392B}. While the spectrum of emitted gravitational waves is likely broad, we note that $\Omega_{\rm BH}$ is in the LISA band for supermassive black holes. Therefore the emission of gravitational waves in this scenario is worth some exploration. The efficiency of the 
black-hole spin energy conversion into the loops is  $\sim (2\pi/\alpha) (R/L)$ and is a strong function of the invariant length $L$ of the loop attached to the black hole. 
Thus determining average $L$ is one of the key targets of future simulations.

\subsection{Primordial black holes}

If both strings and primordial black holes are present in the universe, the strings are formed at a phase transition at a very early time and reach the scaling regime of evolution by the time when black holes are formed.  At that time each horizon volume contains $\mathcal{O}(10)$ long strings and newly formed black holes have sizes comparable to the horizon.  Hence each black hole typically captures $\mathcal{O}(10)$ strings, resulting in an interconnected black hole-string network \cite{2018JCAP...11..008V}.  The strings will wiggle around, cross and reconnect, and it is possible that most of the black holes will be detached from the network.  But even then such black holes will retain string segments with both their ends attached to the horizon.  

Thus, in the primordial black hole scenario we can expect nearly all black holes to end up with string loops attached to them.
The evolution of black hole-string networks is now poorly understood, so we cannot determine the length distribution of the attached string segments.  Progress in this direction would require numerical simulations, which are now being developed. 

Another difference from astrophysical black holes is that primordial black holes are expected to be slowly rotating.  In models where they are formed at high peaks of the density field, their dimensionless angular momentum has been estimated as $\alpha\sim 0.01$ in two independent studies \cite{2020JCAP...03..017M}, \cite{2019JCAP...05..018D}.  This is a small value, but it may be sufficient for the effects described in the preceding sections to be significant.  On the other hand, primordial black holes formed by spherical domain walls or vacuum bubbles nucleated during inflation are non-rotating at birth and can acquire angular momentum only by accretion of matter \cite{2017JCAP...04..050D}, \cite{2017JCAP...12..044D}.  

For black holes with negligible spin, a combination of finite-black-hole mass and black hole spindown will result in an approximately exponential reduction of the length of the loop
attached to the black hole. The loop will disappear after
\begin{equation}
    t\sim 3\times 10^{17}\hbox{s}~M_2\mu_{-21}^{-1}\log\left({L_0\mu^{1/3}\over R}\right),
    \label{lifetime}
\end{equation}
where per common notation, $M_2=M/100M_\odot$, and $L_0$ is the initial size of the loop.
Primordial black holes with loops, if they exist, end up inside galactic halos, but for the loops to survive to the present day, $M$ and $\mu$ must satisfy a rather strict constraint specified by the equation above. This constraint can be relaxed if the black holes can acquire rotational energy through accretion of gas or mergers with other black holes.

	\section{Conclusions}
	The wonderful physics of interactions between black holes and strings, explored  in a long series of publications by Valery Frolov and his collaborators, comes alive if the string is bound into a loop so that its
	both ends are captured by the black hole. The new features that we identify with certainty are 1. the existence of non-self-intersecting loop orbits, 2. the depletion of the
	loop by horizon friction and by reconnections from the secular evolution of the string orbit due to the finite black hole mass, and 3. the growth of the string loop by a superradiant extraction of the black hole's rotational energy. A formalism for the evolution of the
	loop shape has been developed that utilizes a beautiful geometric deformation of an auxiliary curve. It is a matter for future numerical work, to determine the 
	influence of reconnections
	on the sizes  of loops attached to black holes.
	
	We show that  encounters between  string loops and black holes may well be a common occurrence in the real Universe. Gravitational waves could well be an observable signature of such encounters and subsequent evolution of the loops, and in a very general ultra-optimistic scenario their background exceeds the projected sensitivity of LISA by six orders of magnitude. Concrete predictions must wait, however, for future numerical work. 
	
	Finally, we note that there is another potential cause of catastrophic reconnection that we have ignored in this paper. The solution  in Eq.~({\ref{frolov})} describes a string that winds many times around the horizon.  In our case we will have both ends of the loop behave in this way.  There is a danger that these winding ends of the loop will intersect. We suspect that this danger is not so severe, since the spirals are ordered, each occupying its own cone of polar angle $\theta$ and each with a different asymptotic angle $\phi_0$. Thus the two stationary string spirals with different asymptotes never intersect. However the problem deserves a careful consideration once the strings' movement is taken into account.
	
This paper clearly motivates further theoretical studies of the string superradiance, as  well as observational searches for gravitational waves from loops ejected from galactic nuclei.

	We thank Elena Murchikova, Ken Olum, Magdalena Siwek, and David Spergel for useful discussions. The work of AV was supported in part by the National Science Foundation under grant PHY-1820872.
	
	\appendix 
	\renewcommand\thefigure{\thesection.\arabic{figure}}    
	\section{Numerical algorithm for generating random strings}
	\label{app:gen}
	\setcounter{figure}{0}    
	As discussed in Section \ref{sec:num}, we simulate a random string loop by generating a set of vectors ${\bf a}^{\prime}$ which satisfies:
	\begin{enumerate}
		\item $|{\bf a}^{\prime}_i|=1$ (on unit sphere)
		\item $\sum {\bf a}^{\prime}_i=0$ (periodicity)
		\item $|{\bf a}^{\prime}_{i+1}-{\bf a}^{\prime}_i|\le B^{\prime}=2\sin\theta_m $ (smoothness)
	\end{enumerate}
	It is trivial to construct a random vector chain that satisfies conditions 1. and 3. However, in order to satisfy constraint 2., we need to adjust the chain and iterate. An outline of the general procedure is given in Algorithm \ref{alg:1}. This algorithm will guarantee that the $2N$ output vectors  all lie on the unit sphere, that the ${\bf a}(\sigma)$ curve made of equal-length segments parallel to the output vectors is periodic within some tolerance ($|\sum {\bf a}^{\prime}_i|\le \epsilon$) and that it is sufficiently smooth.
	
	\begin{algorithm}[H]
		\caption{Algorithm for generating a random string}
		\label{alg:1}
		\begin{algorithmic}[1]
		\State \textbf{Input}: $N$, $B^{\prime}$, $\epsilon$
		\State \textbf{Output}: A set of 2N vectors ${\bf a}^{\prime}$ that satisfy the all three conditions
		\Do
			\State Generate $2N$ unit vectors with $|{\bf a}^{\prime}_{i+1}-{\bf a}^{\prime}_i| \le f(B^{\prime})$; \label{alg:1:ran}
			\State $\forall i, {\bf a}^{\prime}_i\to {\bf a}^{\prime}_i+(i/2N)(1-f(B)/|{\bf a}^{\prime}_1-{\bf a}^{\prime}_{2N}|)({\bf a}^{\prime}_1-{\bf a}^{\prime}_{2N})$\label{alg:1:chainadj}
			\While {$|\sum {\bf a}^{\prime}_i|>\epsilon$}\label{alg:1:while}
				\State $\forall i, {\bf a}^{\prime}_{i}\to {\bf a}^{\prime}_{i}-\sum {\bf a}^{\prime}_i/2N$;\label{alg:1:per}
				\State $\forall i, {\bf a}^{\prime}_{i}\to {\bf a}^{\prime}_{i} /|{\bf a}^{\prime}_{i}|$;\label{alg:1:norm}
			\EndWhile
		\doWhile {$\max|{\bf a}^{\prime}_{i+1}-{\bf a}^{\prime}_i| > B^{\prime}$}
	\end{algorithmic}
	\end{algorithm}

	At line \ref{alg:1:ran} of Algorithm \ref{alg:1}, we  generate a set of random unit vectors with a random walk step upper bounded by $f(B^{\prime})$. The function mapping $f$ is chosen so that the final vector chain (approximately) satisfies constraint 3. This requires some experimentation and will be discussed later. 
	
	First we choose a unit vector ${\bf a}^{\prime}_1$ in a random direction.  
	We then sequentially generate ${\bf a}^{\prime}_{i+1}$ for each ${\bf a}^{\prime}_{i}$ until we have $2N$ vectors. We orient ${\bf a}^{\prime}_{i}$ as the polar axis and let $\theta$ and $\phi$ be the polar and azimuthal angle the next vector makes with the current one. We draw $\phi$ uniformly in $[0, 2\pi]$. To satisfy the random walk constraint, we  require 
	\begin{equation}
		\theta\le\theta_{m} = 2\arcsin\frac{f(B^{\prime})}{2}.
	\end{equation}
	We choose ${\bf a}^{\prime}_{i+1}$ uniformly from this solid angle, by  drawing $\theta$ from the distribution
	\begin{equation}
		P(\theta) = \frac{\sin\theta}{1-\cos\theta_{max}}.
	\end{equation}
	We use inverse sampling to achieve this. First, we define the cumulative distribution function as 
	\begin{equation}
			Q(\theta) = \int_{0}^{\theta}P(\theta^{\prime})d\theta^{\prime} = \frac{1-\cos\theta}{1-\cos\theta_{m}}\label{cdf}
	\end{equation}
	
	 We can then draw $\theta = Q^{-1}(\chi)$ where $\chi$ is a uniform random variable drawn from $[0, 1]$. We then apply spherical trigonometry to retrieve the Cartesian coordinates of ${\bf a}^{\prime}_{i+1}$.
	 
	 Now, since ${\bf a}$ (and hence ${\bf a}^{\prime}$) is periodic in $2L$, we also need to constrain the starting and ending points of the chain ($|{\bf a}^{\prime}_1-{\bf a}^{\prime}_{2N}|$). Line \ref{alg:1:chainadj} achieves this through shifting each vectors by an amount that is linear in its index. This will ensure the tight inequality in Eq. \ref{eq:newineq}. Here ${\bf \alpha}^{\prime}$ is ${\bf a}^{\prime}$ obtained in Line \ref{alg:1:ran}.
	 \begin{equation}
        |{\bf a}^{\prime}_{(i+1)\text{ mod } 2N}-{\bf a}^{\prime}_{i}|\le \frac{1}{2N}[|{\bf \alpha}^{\prime}_1-{\bf \alpha}^{\prime}_{2N}|+(2N-1)B].\label{eq:newineq}
	 \end{equation}
	
	The next part of the algorithm repeatedly and alternatively makes the vectors periodic (line \ref{alg:1:per}) and normalizes them unto the unit sphere (line \ref{alg:1:norm}). Both conditions will be satisfied (within tolerance), after sufficient iterations. We do not have a rigorous proof of its convergence (and the algorithm may not converge if the initial vectors are linearly dependent). Empirically, the number of iteration required to achieve a tolerance of $10^{-10}$ is plotted in Figure \ref{fig:iter_curve}. We need fewer iterations for larger $N$ and for larger $f(B^{\prime})$. The algorithm will converge in few iterations for reasonable input parameters.
	
	\begin{figure}
	\centering
		\includegraphics[width=\linewidth]{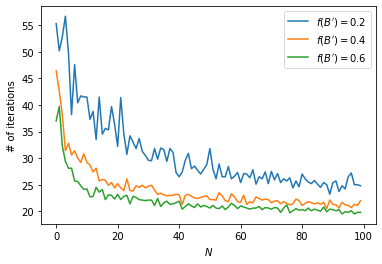}  
		\caption{Iteration required for the algorithm to converge for different values of $N$ and $f(B^{\prime})$ ($\epsilon =10^{-10}$). The results are averaged across 10 independent trials.}
		\label{fig:iter_curve}
	\end{figure}

	Now we use $f(B^{\prime})$ instead of $B^{\prime}$ at line \ref{alg:1:ran} because the iterations do not preserve the maximum random walk step. Empirically, we observe that the maximum step before and after line \ref{alg:1:while}-\ref{alg:1:norm} of the algorithm exhibit a linear relationship (as shown in Figure \ref{fig:f_line}). In practice, to simulate a closed chain with the desired $N$ and $B^{\prime}$, we  first calculate this linear relationship by simulating initial chains with a range of values of $f$ and calculating corresponding values of $B^{\prime}$ once the iterations converge. Using linearity, we  then obtain an estimate for $f$ that we can use for the initial chain in order to obtain a final chain with the
	specified $B^{\prime}$. We thus add the external ``while'' loop to guarantee the correct random walk step bound. We need fewer iterations for a larger number of segments.

	\begin{figure}
		\centering
		\begin{subfigure}{0.49\columnwidth}
			\includegraphics[width=\linewidth]{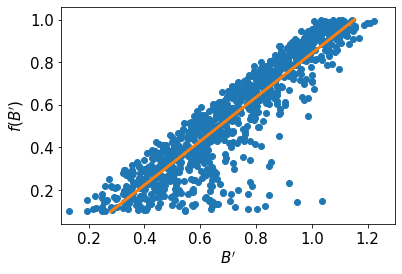}  
			\caption{$N=400$ ($R^2=83\%$)}
			\label{fig:f_line_1}
		\end{subfigure}
		\begin{subfigure}{0.49\columnwidth}
			\includegraphics[width=\linewidth]{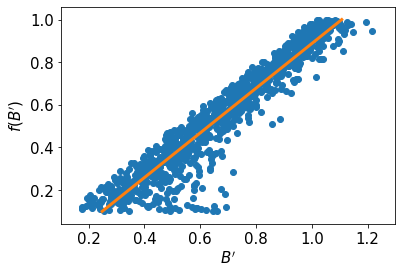}  
			\caption{$N=600$ ($R^2=90\%$)}
			\label{fig:f_line_2}
		\end{subfigure}
		
		\begin{subfigure}{0.49\columnwidth}
			\includegraphics[width=\linewidth]{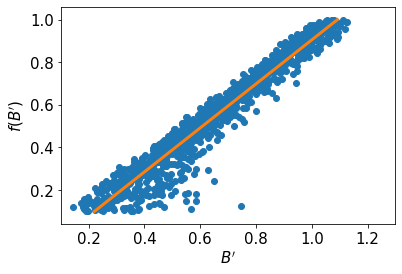}  
			\caption{$N=800$ ($R^2=94\%$)}
			\label{fig:f_line_3}
		\end{subfigure}
	\begin{subfigure}{0.49\columnwidth}
		\includegraphics[width=\linewidth]{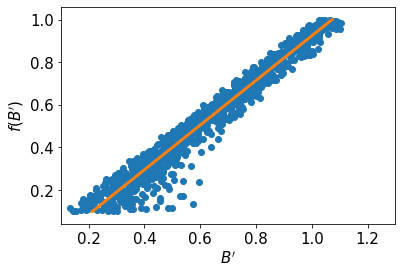}  
		\caption{$N=1000$ ($R^2=96\%$)}
		\label{fig:f_line_4}
	\end{subfigure}
		\caption{Linear relationship between the maximum step size before and after the line \ref{alg:1:while}-\ref{alg:1:norm} of Algorithm \ref{alg:1}.}
		\label{fig:f_line}
	\end{figure}
	
	\section{Numerical algorithm for self-intersections}
	\label{app:det}
	
	Since our loop motion is periodic in $2L$, it suffices for our algorithm to detect intersection with $0\le t<2L$. We divide this time period into intervals of $\Delta t$ and search for intersection within each interval iteratively. Let $t_0$ be the time that an interval starts and $t_0+\Delta t$ is the time it ends. One can choose any value from $0$ to $2L$ for $\Delta t$. In practice we can choose $\Delta t=L/N$ to be the average length of segments which gives satisfactory results. Now, we assume the intersection occurs at location $\sigma_i$ and $\sigma_j$. We have:
	
	\begin{equation}
		{\bf r}(\sigma_i, t) = {\bf r}(\sigma_j, t).
	\end{equation}
	
	With our formulation in Eq. (\ref{gensol}), this becomes:
	
	\begin{equation}
		{\bf a}(\sigma_i-t)-{\bf a}(-\sigma_i-t)={\bf a}(\sigma_j-t)-{\bf a}(-\sigma_j-t). \label{int}
	\end{equation}
	
	Now, as specified in Eq. (\ref{approx}), our algorithm discretizes ${\bf a}$ into $2N$ connected line segments. We assume that $\sigma_i-t$ is on the segment $m_i$, $\sigma_j-t$ is on the segment $m_j$, $-\sigma_i-t$ is on the segment $n_i$ and $-\sigma_j-t$ is on the segment $n_j$. We have:
	\begin{align}
			{\bf a}(\sigma_i-t)=&\sum_{p<m_i}L_p{\bf a}^{\prime}_p+\notag\\&[(\sigma_i-t)\text{ mod }2L-\sum_{p<m_i}L_p]{\bf a}^{\prime}_{m_i},
	\end{align}
	constrained by:
	\begin{equation}
		0\le (\sigma_i-t)\text{ mod }2L - \sum_{p<m_i}L_p<L_{m_i},\label{constraint1}
	\end{equation}
	and similarly for $\sigma_j-t$, $-\sigma_i-t$ and $-\sigma_j-t$. Here mod is real number modulo function defined as $x\text{ mod }y=x-y\lfloor{x/y}\rfloor$. Since ${\bf a}$ is periodic in $2L$, we first use the mod function to make $0\le\sigma_i-t<2L$ where our discretization scheme is defined. To work with mod, we let:
	\begin{align}
		&(\sigma_i-t)\text{ mod }2L = \sigma_i-t + 2M_iL\\
		&(\sigma_j-t)\text{ mod }2L = \sigma_j-t + 2M_jL\\
		&(-\sigma_i-t)\text{ mod }2L = -\sigma_i-t + 2N_iL\\
		&(-\sigma_j-t)\text{ mod }2L = -\sigma_j-t + 2N_jL
	\end{align}
	where $M_i, M_j, N_i, N_j$ are integers. Consequently, Eq. (\ref{int}) becomes:
	\begin{align}
		&({\bf a}^{\prime}_{m_i}+{\bf a}^{\prime}_{n_i}) \sigma_i-({\bf a}^{\prime}_{m_j}+{\bf a}^{\prime}_{n_j}) \sigma_j + \notag\\
		&(-{\bf a}^{\prime}_{m_i}+{\bf a}^{\prime}_{m_j}+{\bf a}^{\prime}_{n_i}-{\bf a}^{\prime}_{n_j})t\notag\\
		=&\sum_{p<m_i}({\bf a}^{\prime}_{m_i}-{\bf a}^{\prime}_{p})L_p-\sum_{p<m_j}({\bf a}^{\prime}_{m_j}-{\bf a}^{\prime}_{p})L_p-\notag\\
		&\sum_{p<n_i}({\bf a}^{\prime}_{n_i}-{\bf a}^{\prime}_{p})L_p+\sum_{p<n_j}({\bf a}^{\prime}_{n_j}-{\bf a}^{\prime}_{p})L_p+\notag\\
		&2L(-M_i{\bf a}^{\prime}_{m_i}+M_j{\bf a}^{\prime}_{m_j}+N_i{\bf a}^{\prime}_{n_i}-N_j{\bf a}^{\prime}_{n_j}).\label{eq:lineq}
	\end{align}
	This is a linear equation system with 3 variables. We look for a solution that satisfies the following constraints:
	
	\begin{equation}
		\begin{cases}
		t_0\le t<t_0+\Delta t\\
		0\le \sigma_i, \sigma_j<L
		\end{cases}\label{ineq1}
	\end{equation}
	
	Furthermore, for our discretization scheme to work, we also need the constraints from Eq. (\ref{constraint1}):
	\begin{equation}
	\begin{cases}
	0\le \sigma_i - t -\sum_{p<m_i}L_p+2M_iL<L_{m_i}\\
	0\le \sigma_j - t -\sum_{p<m_j}L_p+2M_jL<L_{m_j}\\
	0\le -\sigma_i - t -\sum_{p<n_i}L_p+2N_iL<L_{n_i}\\
	0\le -\sigma_j - t -\sum_{p<n_j}L_p+2N_jL<L_{n_j}
	\end{cases}\label{ineq2}
	\end{equation}
	
	Naively, we can try all integer combinations of $M_i, M_j, N_i, N_j, m_i, m_j, n_i, n_j$ and check if Eq. (\ref{eq:lineq}) has a solution. However, this algorithm will then have time complexity $\mathcal{O}(N^8L/\Delta t)$ and is not computationally feasible. We thus seek to constrain these parameters and trim our search space. 
	
	\subsection{Constraining $M_i, M_j, N_i, N_j$}
	
	From the inequalities in (\ref{ineq2}), we have:
	\begin{align}
		&\sum\limits_{p<m_i}L_p-\sigma_i+t\le 2M_iL<\sum\limits_{p\le m_i}L_p-\sigma_i+t\\
		&\sum\limits_{p<n_i}L_p+\sigma_i+t\le 2N_iL<\sum\limits_{p\le n_i}L_p-\sigma_i+t
	\end{align}
	and similarly for $M_j$ and $N_j$. Then, using the inequalities in (\ref{ineq1}) as well as $0\le\sum_{p<k}L_p\le L$ for all $k$, we obtain the following constraints:
	\begin{align}
		&(t_0-L)/2L\le M_i, M_j<(t_0+L+\Delta t)/2L\\
		&t_0/2L\le M_i, M_j<(t_0+2L+\Delta t)/2L
	\end{align}
	Consequently, there are at most 2 feasible values for each of $M_i$, $M_j$, $N_i$ and $N_j$ (since $\Delta t < 2L$). We are left with at most 16 combinations to try, effectively reducing the time complexity to $\mathcal{O}(N^4L/\Delta t)$. 
	
	\subsection{Constraining $n_i, n_j$}
	
	We can further accelerate the algorithm by constraining $n_i$ and $n_j$ given $m_i$, $m_j$, $M_i$, $M_j$, $N_i$ and $N_j$. Using the inequalities in (\ref{ineq2}) and $t<t_0+2L$, we have:
	\begin{align}
		& \sum_{p< n_i}L_p\le-\sum_{p< m_i}L_p-2t_0+2(M_i+N_i)L\label{consn1}\\
		& \sum_{p< n_j}L_p\le-\sum_{p< m_j}L_p-2t_0+2(M_j+N_j)L\\
		& \sum_{p\le n_i}L_p>-\sum_{p\le m_i}L_p-2t_0-2\Delta t+2(M_i+N_i)L\label{consn2}\\
		& \sum_{p\le n_j}L_p>-\sum_{p\le m_j}L_p-2t_0-2\Delta t+2(M_j+N_j)L
	\end{align}
	
	These will filter out a significant proportion of the available values for $n_i$ and $n_j$. The right hand side of inequalities (\ref{consn1}) and (\ref{consn2}) differ by $L_{m_i}+2\Delta t$. If there are $k$ segments falling into this range, we will have $k+2$ possible values for $n_i$. Assigning the average segment length to $\Delta t$, we will have $k\sim 3$. Thus, on average, we will have approximately 5 $n_i$ values to try. This reduces the time complexity to $\mathcal{O}(N^2L/\Delta t)=\mathcal{O}(N^3)$.
	
	\subsection{Constraining $m_j$}
	
	Moreover, with a given $m_i$, we can develop a constraint on the available values for $m_j$. From the inequalities in (\ref{ineq2}), we can find a bound for $\sigma_i$ and $\sigma_j$:
	\begin{align}
		&t_0+\sum_{p< m_i}L_p-2M_iL\le \sigma_i<t_0+\Delta t+\sum_{p\le m_i}L_p-2M_iL\\
		&t_0+\sum_{p< m_j}L_p-2M_jL\le \sigma_j<t_0+\Delta t+\sum_{p\le m_j}L_p-2M_jL
	\end{align}
	Note that the right side of these inequalities are larger than the left side by $\Delta t+L_{m_i}$ and $\Delta t+L_{m_j}$, respectively. That is, the $\sigma$ values are constrained within this region given $m_i$ and $m_j$. Now, at $t_0$, the spatial distance between two intersection points has:
	\begin{align}
		dist =&||{\bf r}(\sigma_i, t_0) - {\bf r}(\sigma_j, t_0)||\notag\\
		\ge& ||{\bf r}(t_0+\sum_{p< m_i}L_p-2M_iL, t_0) -\notag\\
		& {\bf r}(t_0+\sum_{p< m_j}L_p-2M_jL, t_0)||-\notag\\
		&2\Delta t - L_{m_i} - L_{m_j}
	\end{align}
	The last three terms come from the aforementioned two intervals for $\sigma$. When varying $\sigma$, the two points can approach each other by at most $2 \Delta t+ L_{m_i} + L_{m_j}$. Then, in the time interval $\Delta t$, these two points will move towards each other to form an intersection. Both points move at a maximum speed of 1 (light speed) so the distance must satisfy:
	
	\begin{equation}
		dist < 2\Delta t
	\end{equation}
	 This will give us the constraint on $m_j$:
	 \begin{align}
	 	||&{\bf r}(t_0+\sum_{p< m_i}L_p-2M_iL, t_0) -\notag\\
	 	& {\bf r}(t_0+\sum_{p< m_j}L_p-2M_jL, t_0)||\le 4\Delta t + L_{m_i} + L_{m_j}
	 \end{align}
	 
	 The effectiveness of this constraint will rely on the shape of the string. For example, if the string is a perfect circle ($r=L/2\pi$), the number of $m_j$ values that satisfy this constraint is approximately
	 
	 \begin{equation}
	 	\frac{2\arcsin((2\Delta t+L/N)/r)}{2\pi}N\approx 6
	 \end{equation}
	 
	 assuming $N\gg 1$. In general, the smoother the string is (or the smaller $B^{\prime}$ is), the more values that we can filter out.
	 
	 \subsection{Computational imprecision}
	 
	 Our scheme of using segments is not a precise model of a smooth physical string. Particularly, if $\sigma_i$ and $\sigma_j$ are too close, the discrete nature of our string will be manifest. So the detected intersection could simply be due to the failure of our approximation. Therefore, we filter out the intersections that are too close together by requiring:
	 
	 \begin{equation}
	 	\min\{|\sigma_i-\sigma_j|, |\sigma_i-\sigma_j+L|\}>\text{threshold}
	 \end{equation}
	 
	 Empirically, we find that the algorithm is stable if we choose $\text{threshold}=10L/N$. That is, if two intersections are separated by, on average, at least 10 segments, we are confident that this is not due to numerical error.
	 
	 This also means that if our string length becomes smaller than $10L_i/N$ where $L_i$ is the initial length, our algorithm is no longer entirely reliable. As shown in Figure \ref{fig:int_line}, we can see that the final lengths of our simulated string are well above this threshold ($2\times10^{-3}$).

	\bibliography{main}

\begin{thebibliography}{45}
\expandafter\ifx\csname natexlab\endcsname\relax\def\natexlab#1{#1}\fi
\expandafter\ifx\csname bibnamefont\endcsname\relax
  \def\bibnamefont#1{#1}\fi
\expandafter\ifx\csname bibfnamefont\endcsname\relax
  \def\bibfnamefont#1{#1}\fi
\expandafter\ifx\csname citenamefont\endcsname\relax
  \def\citenamefont#1{#1}\fi
\expandafter\ifx\csname url\endcsname\relax
  \def\url#1{\texttt{#1}}\fi
\expandafter\ifx\csname urlprefix\endcsname\relax\def\urlprefix{URL }\fi
\providecommand{\bibinfo}[2]{#2}
\providecommand{\eprint}[2][]{\url{#2}}

\bibitem[{\citenamefont{{Kibble}}(1976)}]{1976JPhA....9.1387K}
\bibinfo{author}{\bibfnamefont{T.~W.~B.} \bibnamefont{{Kibble}}},
  \bibinfo{journal}{Journal of Physics A Mathematical General}
  \textbf{\bibinfo{volume}{9}}, \bibinfo{pages}{1387} (\bibinfo{year}{1976}).

\bibitem[{\citenamefont{{Vilenkin} and {Shellard}}(2000)}]{2000csot.book.....V}
\bibinfo{author}{\bibfnamefont{A.}~\bibnamefont{{Vilenkin}}} \bibnamefont{and}
  \bibinfo{author}{\bibfnamefont{E.~P.~S.} \bibnamefont{{Shellard}}},
  \emph{\bibinfo{title}{{Cosmic Strings and Other Topological Defects}}}
  (\bibinfo{year}{2000}).

\bibitem[{\citenamefont{{Jones} et~al.}(2003)\citenamefont{{Jones}, {Stoica},
  and {Tye}}}]{2003PhLB..563....6J}
\bibinfo{author}{\bibfnamefont{N.~T.} \bibnamefont{{Jones}}},
  \bibinfo{author}{\bibfnamefont{H.}~\bibnamefont{{Stoica}}}, \bibnamefont{and}
  \bibinfo{author}{\bibfnamefont{S.~H.~H.} \bibnamefont{{Tye}}},
  \bibinfo{journal}{Physics Letters B} \textbf{\bibinfo{volume}{563}},
  \bibinfo{pages}{6} (\bibinfo{year}{2003}), \eprint{hep-th/0303269}.

\bibitem[{\citenamefont{{Vilenkin}}(1981)}]{1981PhLB..107...47V}
\bibinfo{author}{\bibfnamefont{A.}~\bibnamefont{{Vilenkin}}},
  \bibinfo{journal}{Physics Letters B} \textbf{\bibinfo{volume}{107}},
  \bibinfo{pages}{47} (\bibinfo{year}{1981}).

\bibitem[{\citenamefont{{Damour} and {Vilenkin}}(2005)}]{2005PhRvD..71f3510D}
\bibinfo{author}{\bibfnamefont{T.}~\bibnamefont{{Damour}}} \bibnamefont{and}
  \bibinfo{author}{\bibfnamefont{A.}~\bibnamefont{{Vilenkin}}},
  \bibinfo{journal}{\prd} \textbf{\bibinfo{volume}{71}}, \bibinfo{eid}{063510}
  (\bibinfo{year}{2005}), \eprint{hep-th/0410222}.

\bibitem[{\citenamefont{{Vilenkin} et~al.}(2018)\citenamefont{{Vilenkin},
  {Levin}, and {Gruzinov}}}]{2018JCAP...11..008V}
\bibinfo{author}{\bibfnamefont{A.}~\bibnamefont{{Vilenkin}}},
  \bibinfo{author}{\bibfnamefont{Y.}~\bibnamefont{{Levin}}}, \bibnamefont{and}
  \bibinfo{author}{\bibfnamefont{A.}~\bibnamefont{{Gruzinov}}},
  \bibinfo{journal}{Journal of Cosmology and Astroparticle Physics}
  \textbf{\bibinfo{volume}{2018}}, \bibinfo{eid}{008} (\bibinfo{year}{2018}),
  \eprint{1808.00670}.

\bibitem[{\citenamefont{{Lonsdale} and {Moss}}(1988)}]{1988NuPhB.298..693L}
\bibinfo{author}{\bibfnamefont{S.}~\bibnamefont{{Lonsdale}}} \bibnamefont{and}
  \bibinfo{author}{\bibfnamefont{I.}~\bibnamefont{{Moss}}},
  \bibinfo{journal}{Nuclear Physics B} \textbf{\bibinfo{volume}{298}},
  \bibinfo{pages}{693} (\bibinfo{year}{1988}).

\bibitem[{\citenamefont{{de Villiers} and
  {Frolov}}(1998)}]{1998IJMPD...7..957D}
\bibinfo{author}{\bibfnamefont{J.-P.} \bibnamefont{{de Villiers}}}
  \bibnamefont{and} \bibinfo{author}{\bibfnamefont{V.}~\bibnamefont{{Frolov}}},
  \bibinfo{journal}{International Journal of Modern Physics D}
  \textbf{\bibinfo{volume}{7}}, \bibinfo{pages}{957} (\bibinfo{year}{1998}),
  \eprint{gr-qc/9711045}.

\bibitem[{\citenamefont{{DeVilliers} and {Frolov}}(1999)}]{1999CQGra..16.2403D}
\bibinfo{author}{\bibfnamefont{J.-P.} \bibnamefont{{DeVilliers}}}
  \bibnamefont{and} \bibinfo{author}{\bibfnamefont{V.}~\bibnamefont{{Frolov}}},
  \bibinfo{journal}{Classical and Quantum Gravity}
  \textbf{\bibinfo{volume}{16}}, \bibinfo{pages}{2403} (\bibinfo{year}{1999}),
  \eprint{gr-qc/9812016}.

\bibitem[{\citenamefont{{Snajdr} and {Frolov}}(2003)}]{2003CQGra..20.1303S}
\bibinfo{author}{\bibfnamefont{M.}~\bibnamefont{{Snajdr}}} \bibnamefont{and}
  \bibinfo{author}{\bibfnamefont{V.}~\bibnamefont{{Frolov}}},
  \bibinfo{journal}{Classical and Quantum Gravity}
  \textbf{\bibinfo{volume}{20}}, \bibinfo{pages}{1303} (\bibinfo{year}{2003}),
  \eprint{gr-qc/0211018}.

\bibitem[{\citenamefont{{Dubath} et~al.}(2007)\citenamefont{{Dubath},
  {Sakellariadou}, and {Viallet}}}]{2007IJMPD..16.1311D}
\bibinfo{author}{\bibfnamefont{F.}~\bibnamefont{{Dubath}}},
  \bibinfo{author}{\bibfnamefont{M.}~\bibnamefont{{Sakellariadou}}},
  \bibnamefont{and} \bibinfo{author}{\bibfnamefont{C.~M.}
  \bibnamefont{{Viallet}}}, \bibinfo{journal}{International Journal of Modern
  Physics D} \textbf{\bibinfo{volume}{16}}, \bibinfo{pages}{1311}
  (\bibinfo{year}{2007}), \eprint{gr-qc/0609089}.

\bibitem[{\citenamefont{{Frolov} et~al.}(1996)\citenamefont{{Frolov}, {Hendy},
  and {Larsen}}}]{1996PhRvD..54.5093F}
\bibinfo{author}{\bibfnamefont{V.}~\bibnamefont{{Frolov}}},
  \bibinfo{author}{\bibfnamefont{S.}~\bibnamefont{{Hendy}}}, \bibnamefont{and}
  \bibinfo{author}{\bibfnamefont{A.~L.} \bibnamefont{{Larsen}}},
  \bibinfo{journal}{\prd} \textbf{\bibinfo{volume}{54}}, \bibinfo{pages}{5093}
  (\bibinfo{year}{1996}), \eprint{hep-th/9510231}.

\bibitem[{\citenamefont{{Igata} et~al.}(2018)\citenamefont{{Igata}, {Ishihara},
  {Tsuchiya}, and {Yoo}}}]{2018PhRvD..98f4021I}
\bibinfo{author}{\bibfnamefont{T.}~\bibnamefont{{Igata}}},
  \bibinfo{author}{\bibfnamefont{H.}~\bibnamefont{{Ishihara}}},
  \bibinfo{author}{\bibfnamefont{M.}~\bibnamefont{{Tsuchiya}}},
  \bibnamefont{and} \bibinfo{author}{\bibfnamefont{C.-M.} \bibnamefont{{Yoo}}},
  \bibinfo{journal}{\prd} \textbf{\bibinfo{volume}{98}}, \bibinfo{eid}{064021}
  (\bibinfo{year}{2018}), \eprint{1806.09837}.

\bibitem[{\citenamefont{{Blanco-Pillado}
  et~al.}(2018)\citenamefont{{Blanco-Pillado}, {Olum}, and
  {Siemens}}}]{2018PhLB..778..392B}
\bibinfo{author}{\bibfnamefont{J.~J.} \bibnamefont{{Blanco-Pillado}}},
  \bibinfo{author}{\bibfnamefont{K.~D.} \bibnamefont{{Olum}}},
  \bibnamefont{and}
  \bibinfo{author}{\bibfnamefont{X.}~\bibnamefont{{Siemens}}},
  \bibinfo{journal}{Physics Letters B} \textbf{\bibinfo{volume}{778}},
  \bibinfo{pages}{392} (\bibinfo{year}{2018}), \eprint{1709.02434}.

\bibitem[{\citenamefont{{Ach{\'u}carro} and {de
  Putter}}(2006)}]{2006PhRvD..74l1701A}
\bibinfo{author}{\bibfnamefont{A.}~\bibnamefont{{Ach{\'u}carro}}}
  \bibnamefont{and} \bibinfo{author}{\bibfnamefont{R.}~\bibnamefont{{de
  Putter}}}, \bibinfo{journal}{\prd} \textbf{\bibinfo{volume}{74}},
  \bibinfo{eid}{121701} (\bibinfo{year}{2006}), \eprint{hep-th/0605084}.

\bibitem[{\citenamefont{{Jackson} et~al.}(2005)\citenamefont{{Jackson},
  {Jones}, and {Polchinski}}}]{2005JHEP...10..013J}
\bibinfo{author}{\bibfnamefont{M.~G.} \bibnamefont{{Jackson}}},
  \bibinfo{author}{\bibfnamefont{N.~T.} \bibnamefont{{Jones}}},
  \bibnamefont{and}
  \bibinfo{author}{\bibfnamefont{J.}~\bibnamefont{{Polchinski}}},
  \bibinfo{journal}{Journal of High Energy Physics}
  \textbf{\bibinfo{volume}{2005}}, \bibinfo{eid}{013} (\bibinfo{year}{2005}),
  \eprint{hep-th/0405229}.

\bibitem[{\citenamefont{York}(1989)}]{PhysRevD.40.277}
\bibinfo{author}{\bibfnamefont{T.}~\bibnamefont{York}}, \bibinfo{journal}{Phys.
  Rev. D} \textbf{\bibinfo{volume}{40}}, \bibinfo{pages}{277}
  (\bibinfo{year}{1989}).

\bibitem[{\citenamefont{{Siemens} et~al.}(2001)\citenamefont{{Siemens},
  {Martin}, and {Olum}}}]{2001NuPhB.595..402S}
\bibinfo{author}{\bibfnamefont{X.}~\bibnamefont{{Siemens}}},
  \bibinfo{author}{\bibfnamefont{X.}~\bibnamefont{{Martin}}}, \bibnamefont{and}
  \bibinfo{author}{\bibfnamefont{K.~D.} \bibnamefont{{Olum}}},
  \bibinfo{journal}{Nuclear Physics B} \textbf{\bibinfo{volume}{595}},
  \bibinfo{pages}{402} (\bibinfo{year}{2001}), \eprint{astro-ph/0005411}.

\bibitem[{\citenamefont{{Frolov} et~al.}(1989)\citenamefont{{Frolov},
  {Skarzhinsky}, {Zelnikov}, and {Heinrich}}}]{1989PhLB..224..255F}
\bibinfo{author}{\bibfnamefont{V.~P.} \bibnamefont{{Frolov}}},
  \bibinfo{author}{\bibfnamefont{V.~D.} \bibnamefont{{Skarzhinsky}}},
  \bibinfo{author}{\bibfnamefont{A.~I.} \bibnamefont{{Zelnikov}}},
  \bibnamefont{and}
  \bibinfo{author}{\bibfnamefont{O.}~\bibnamefont{{Heinrich}}},
  \bibinfo{journal}{Physics Letters B} \textbf{\bibinfo{volume}{224}},
  \bibinfo{pages}{255} (\bibinfo{year}{1989}).

\bibitem[{\citenamefont{{Boos} and {Frolov}}(2018)}]{2018PhRvD..97b4024B}
\bibinfo{author}{\bibfnamefont{J.}~\bibnamefont{{Boos}}} \bibnamefont{and}
  \bibinfo{author}{\bibfnamefont{V.~P.} \bibnamefont{{Frolov}}},
  \bibinfo{journal}{\prd} \textbf{\bibinfo{volume}{97}}, \bibinfo{eid}{024024}
  (\bibinfo{year}{2018}), \eprint{1711.06357}.

\bibitem[{\citenamefont{{Hawking} and {Hartle}}(1972)}]{1972CMaPh..27..283H}
\bibinfo{author}{\bibfnamefont{S.~W.} \bibnamefont{{Hawking}}}
  \bibnamefont{and} \bibinfo{author}{\bibfnamefont{J.~B.}
  \bibnamefont{{Hartle}}}, \bibinfo{journal}{Communications in Mathematical
  Physics} \textbf{\bibinfo{volume}{27}}, \bibinfo{pages}{283}
  (\bibinfo{year}{1972}).

\bibitem[{\citenamefont{{Zel'Dovich}}(1971)}]{1971JETPL..14..180Z}
\bibinfo{author}{\bibfnamefont{Y.~B.} \bibnamefont{{Zel'Dovich}}},
  \bibinfo{journal}{Soviet Journal of Experimental and Theoretical Physics
  Letters} \textbf{\bibinfo{volume}{14}}, \bibinfo{pages}{180}
  (\bibinfo{year}{1971}).

\bibitem[{\citenamefont{{Press} and {Teukolsky}}(1972)}]{1972Natur.238..211P}
\bibinfo{author}{\bibfnamefont{W.~H.} \bibnamefont{{Press}}} \bibnamefont{and}
  \bibinfo{author}{\bibfnamefont{S.~A.} \bibnamefont{{Teukolsky}}},
  \bibinfo{journal}{\nat} \textbf{\bibinfo{volume}{238}}, \bibinfo{pages}{211}
  (\bibinfo{year}{1972}).

\bibitem[{\citenamefont{{Gage}}(1984)}]{1984InMat..76..357G}
\bibinfo{author}{\bibfnamefont{M.~E.} \bibnamefont{{Gage}}},
  \bibinfo{journal}{Inventiones Mathematicae} \textbf{\bibinfo{volume}{76}},
  \bibinfo{pages}{357} (\bibinfo{year}{1984}).

\bibitem[{\citenamefont{Gage and Hamilton}(1986)}]{gage1986}
\bibinfo{author}{\bibfnamefont{M.}~\bibnamefont{Gage}} \bibnamefont{and}
  \bibinfo{author}{\bibfnamefont{R.~S.} \bibnamefont{Hamilton}},
  \bibinfo{journal}{J. Differential Geom.} \textbf{\bibinfo{volume}{23}},
  \bibinfo{pages}{69} (\bibinfo{year}{1986}),
  \urlprefix\url{https://doi.org/10.4310/jdg/1214439902}.

\bibitem[{\citenamefont{Grayson}(1989)}]{10.2307/1971486}
\bibinfo{author}{\bibfnamefont{M.~A.} \bibnamefont{Grayson}},
  \bibinfo{journal}{Annals of Mathematics} \textbf{\bibinfo{volume}{129}},
  \bibinfo{pages}{71} (\bibinfo{year}{1989}), ISSN \bibinfo{issn}{0003486X},
  \urlprefix\url{http://www.jstor.org/stable/1971486}.

\bibitem[{\citenamefont{{Martin} and {Vilenkin}}(1997)}]{1997PhRvD..55.6054M}
\bibinfo{author}{\bibfnamefont{X.}~\bibnamefont{{Martin}}} \bibnamefont{and}
  \bibinfo{author}{\bibfnamefont{A.}~\bibnamefont{{Vilenkin}}},
  \bibinfo{journal}{\prd} \textbf{\bibinfo{volume}{55}}, \bibinfo{pages}{6054}
  (\bibinfo{year}{1997}), \eprint{gr-qc/9612008}.

\bibitem[{\citenamefont{{Blanco-Pillado} and
  {Olum}}(2020)}]{2020PhRvD.101j3018B}
\bibinfo{author}{\bibfnamefont{J.~J.} \bibnamefont{{Blanco-Pillado}}}
  \bibnamefont{and} \bibinfo{author}{\bibfnamefont{K.~D.}
  \bibnamefont{{Olum}}}, \bibinfo{journal}{\prd}
  \textbf{\bibinfo{volume}{101}}, \bibinfo{eid}{103018} (\bibinfo{year}{2020}),
  \eprint{1912.10017}.

\bibitem[{\citenamefont{{Chernoff}}(2009)}]{2009arXiv0908.4077C}
\bibinfo{author}{\bibfnamefont{D.~F.} \bibnamefont{{Chernoff}}},
  \bibinfo{journal}{arXiv e-prints} \bibinfo{eid}{arXiv:0908.4077}
  (\bibinfo{year}{2009}), \eprint{0908.4077}.

\bibitem[{\citenamefont{{Jain} and {Vilenkin}}(2020)}]{2020arXiv200615358J}
\bibinfo{author}{\bibfnamefont{M.}~\bibnamefont{{Jain}}} \bibnamefont{and}
  \bibinfo{author}{\bibfnamefont{A.}~\bibnamefont{{Vilenkin}}},
  \bibinfo{journal}{arXiv e-prints} \bibinfo{eid}{arXiv:2006.15358}
  (\bibinfo{year}{2020}), \eprint{2006.15358}.

\bibitem[{\citenamefont{{Avgoustidis} and
  {Shellard}}(2006)}]{2006PhRvD..73d1301A}
\bibinfo{author}{\bibfnamefont{A.}~\bibnamefont{{Avgoustidis}}}
  \bibnamefont{and} \bibinfo{author}{\bibfnamefont{E.~P.~S.}
  \bibnamefont{{Shellard}}}, \bibinfo{journal}{\prd}
  \textbf{\bibinfo{volume}{73}}, \bibinfo{eid}{041301} (\bibinfo{year}{2006}),
  \eprint{astro-ph/0512582}.

\bibitem[{\citenamefont{{Young}}(1980)}]{1980ApJ...242.1232Y}
\bibinfo{author}{\bibfnamefont{P.}~\bibnamefont{{Young}}},
  \bibinfo{journal}{\apj} \textbf{\bibinfo{volume}{242}}, \bibinfo{pages}{1232}
  (\bibinfo{year}{1980}).

\bibitem[{\citenamefont{{Shankar} et~al.}(2009)\citenamefont{{Shankar},
  {Weinberg}, and {Miralda-Escud{\'e}}}}]{2009ApJ...690...20S}
\bibinfo{author}{\bibfnamefont{F.}~\bibnamefont{{Shankar}}},
  \bibinfo{author}{\bibfnamefont{D.~H.} \bibnamefont{{Weinberg}}},
  \bibnamefont{and}
  \bibinfo{author}{\bibfnamefont{J.}~\bibnamefont{{Miralda-Escud{\'e}}}},
  \bibinfo{journal}{\apj} \textbf{\bibinfo{volume}{690}}, \bibinfo{pages}{20}
  (\bibinfo{year}{2009}), \eprint{0710.4488}.

\bibitem[{\citenamefont{{Kelly} and {Merloni}}(2012)}]{2012AdAst2012E...7K}
\bibinfo{author}{\bibfnamefont{B.~C.} \bibnamefont{{Kelly}}} \bibnamefont{and}
  \bibinfo{author}{\bibfnamefont{A.}~\bibnamefont{{Merloni}}},
  \bibinfo{journal}{Advances in Astronomy} \textbf{\bibinfo{volume}{2012}},
  \bibinfo{eid}{970858} (\bibinfo{year}{2012}), \eprint{1112.1430}.

\bibitem[{\citenamefont{{Jones} et~al.}(2020)\citenamefont{{Jones},
  {Brenneman}, {Civano}, {Lanzuisi}, and {Marchesi}}}]{2020arXiv200808588J}
\bibinfo{author}{\bibfnamefont{M.}~\bibnamefont{{Jones}}},
  \bibinfo{author}{\bibfnamefont{L.}~\bibnamefont{{Brenneman}}},
  \bibinfo{author}{\bibfnamefont{F.}~\bibnamefont{{Civano}}},
  \bibinfo{author}{\bibfnamefont{G.}~\bibnamefont{{Lanzuisi}}},
  \bibnamefont{and}
  \bibinfo{author}{\bibfnamefont{S.}~\bibnamefont{{Marchesi}}},
  \bibinfo{journal}{arXiv e-prints} \bibinfo{eid}{arXiv:2008.08588}
  (\bibinfo{year}{2020}), \eprint{2008.08588}.

\bibitem[{\citenamefont{{Daly}}(2019)}]{2019ApJ...886...37D}
\bibinfo{author}{\bibfnamefont{R.~A.} \bibnamefont{{Daly}}},
  \bibinfo{journal}{\apj} \textbf{\bibinfo{volume}{886}}, \bibinfo{eid}{37}
  (\bibinfo{year}{2019}), \eprint{1905.11319}.

\bibitem[{\citenamefont{{Fragione} and {Loeb}}(2020)}]{2020arXiv200811734F}
\bibinfo{author}{\bibfnamefont{G.}~\bibnamefont{{Fragione}}} \bibnamefont{and}
  \bibinfo{author}{\bibfnamefont{A.}~\bibnamefont{{Loeb}}},
  \bibinfo{journal}{arXiv e-prints} \bibinfo{eid}{arXiv:2008.11734}
  (\bibinfo{year}{2020}), \eprint{2008.11734}.

\bibitem[{\citenamefont{{Ali} et~al.}(2020)\citenamefont{{Ali}, {Paul},
  {Eckart}, {Parsa}, {Zajacek}, {Pei{\ss}ker}, {Subroweit}, {Valencia-S.},
  {Thomkins}, and {Witzel}}}]{2020ApJ...896..100A}
\bibinfo{author}{\bibfnamefont{B.}~\bibnamefont{{Ali}}},
  \bibinfo{author}{\bibfnamefont{D.}~\bibnamefont{{Paul}}},
  \bibinfo{author}{\bibfnamefont{A.}~\bibnamefont{{Eckart}}},
  \bibinfo{author}{\bibfnamefont{M.}~\bibnamefont{{Parsa}}},
  \bibinfo{author}{\bibfnamefont{M.}~\bibnamefont{{Zajacek}}},
  \bibinfo{author}{\bibfnamefont{F.}~\bibnamefont{{Pei{\ss}ker}}},
  \bibinfo{author}{\bibfnamefont{M.}~\bibnamefont{{Subroweit}}},
  \bibinfo{author}{\bibfnamefont{M.}~\bibnamefont{{Valencia-S.}}},
  \bibinfo{author}{\bibfnamefont{L.}~\bibnamefont{{Thomkins}}},
  \bibnamefont{and} \bibinfo{author}{\bibfnamefont{G.}~\bibnamefont{{Witzel}}},
  \bibinfo{journal}{\apj} \textbf{\bibinfo{volume}{896}}, \bibinfo{eid}{100}
  (\bibinfo{year}{2020}), \eprint{2006.11399}.

\bibitem[{\citenamefont{{Levin} and {Beloborodov}}(2003)}]{2003ApJ...590L..33L}
\bibinfo{author}{\bibfnamefont{Y.}~\bibnamefont{{Levin}}} \bibnamefont{and}
  \bibinfo{author}{\bibfnamefont{A.~M.} \bibnamefont{{Beloborodov}}},
  \bibinfo{journal}{\apj Letters} \textbf{\bibinfo{volume}{590}},
  \bibinfo{pages}{L33} (\bibinfo{year}{2003}), \eprint{astro-ph/0303436}.

\bibitem[{\citenamefont{{Soltan}}(1982)}]{1982MNRAS.200..115S}
\bibinfo{author}{\bibfnamefont{A.}~\bibnamefont{{Soltan}}},
  \bibinfo{journal}{Monthly Notices of the Royal Astronomical Society}
  \textbf{\bibinfo{volume}{200}}, \bibinfo{pages}{115} (\bibinfo{year}{1982}).

\bibitem[{\citenamefont{{Yu} and {Tremaine}}(2002)}]{2002MNRAS.335..965Y}
\bibinfo{author}{\bibfnamefont{Q.}~\bibnamefont{{Yu}}} \bibnamefont{and}
  \bibinfo{author}{\bibfnamefont{S.}~\bibnamefont{{Tremaine}}},
  \bibinfo{journal}{Monthly Notices of the Royal Astronomical Society}
  \textbf{\bibinfo{volume}{335}}, \bibinfo{pages}{965} (\bibinfo{year}{2002}),
  \eprint{astro-ph/0203082}.

\bibitem[{\citenamefont{{Mirbabayi} et~al.}(2020)\citenamefont{{Mirbabayi},
  {Gruzinov}, and {Nore{\~n}a}}}]{2020JCAP...03..017M}
\bibinfo{author}{\bibfnamefont{M.}~\bibnamefont{{Mirbabayi}}},
  \bibinfo{author}{\bibfnamefont{A.}~\bibnamefont{{Gruzinov}}},
  \bibnamefont{and}
  \bibinfo{author}{\bibfnamefont{J.}~\bibnamefont{{Nore{\~n}a}}},
  \bibinfo{journal}{Journal of Cosmology and Astroparticle Physics}
  \textbf{\bibinfo{volume}{2020}}, \bibinfo{eid}{017} (\bibinfo{year}{2020}),
  \eprint{1901.05963}.

\bibitem[{\citenamefont{{De Luca} et~al.}(2019)\citenamefont{{De Luca},
  {Desjacques}, {Franciolini}, {Malhotra}, and {Riotto}}}]{2019JCAP...05..018D}
\bibinfo{author}{\bibfnamefont{V.}~\bibnamefont{{De Luca}}},
  \bibinfo{author}{\bibfnamefont{V.}~\bibnamefont{{Desjacques}}},
  \bibinfo{author}{\bibfnamefont{G.}~\bibnamefont{{Franciolini}}},
  \bibinfo{author}{\bibfnamefont{A.}~\bibnamefont{{Malhotra}}},
  \bibnamefont{and} \bibinfo{author}{\bibfnamefont{A.}~\bibnamefont{{Riotto}}},
  \bibinfo{journal}{Journal of Cosmology and Astroparticle Physics}
  \textbf{\bibinfo{volume}{2019}}, \bibinfo{eid}{018} (\bibinfo{year}{2019}),
  \eprint{1903.01179}.

\bibitem[{\citenamefont{{Deng} et~al.}(2017)\citenamefont{{Deng}, {Garriga},
  and {Vilenkin}}}]{2017JCAP...04..050D}
\bibinfo{author}{\bibfnamefont{H.}~\bibnamefont{{Deng}}},
  \bibinfo{author}{\bibfnamefont{J.}~\bibnamefont{{Garriga}}},
  \bibnamefont{and}
  \bibinfo{author}{\bibfnamefont{A.}~\bibnamefont{{Vilenkin}}},
  \bibinfo{journal}{Journal of Cosmology and Astroparticle Physics}
  \textbf{\bibinfo{volume}{2017}}, \bibinfo{eid}{050} (\bibinfo{year}{2017}),
  \eprint{1612.03753}.

\bibitem[{\citenamefont{{Deng} and {Vilenkin}}(2017)}]{2017JCAP...12..044D}
\bibinfo{author}{\bibfnamefont{H.}~\bibnamefont{{Deng}}} \bibnamefont{and}
  \bibinfo{author}{\bibfnamefont{A.}~\bibnamefont{{Vilenkin}}},
  \bibinfo{journal}{Journal of Cosmology and Astroparticle Physics}
  \textbf{\bibinfo{volume}{2017}}, \bibinfo{eid}{044} (\bibinfo{year}{2017}),
  \eprint{1710.02865}.

\end{thebibliography}
	
	
	
	

	

\end{document}